\documentclass[epj]{svjourmod}

\usepackage{amsmath,amssymb}
\usepackage{graphicx,epsfig}
\usepackage{citesort}

\allowdisplaybreaks[1]

\newcommand{\sh}{\not\!}

\newcommand{\ol}{\overline}

\newcommand{\rR}{R}
\newcommand{\rr}{{\cal R}}
\newcommand{\mc}{\multicolumn{2}{c}}
\newcommand{\mcc}{\multicolumn{3}{c}}

\newcommand{\nl}{\!\!\! & \!\!\!}
\newcommand{\av}[1]{\langle {#1}\rangle}

\def\beq{\begin{equation}}
\def\eeq{\end{equation}}
\def\bea{\begin{eqnarray}}
\def\eea{\end{eqnarray}}
\newcommand{\bes}{\begin{split}}
\newcommand{\ees}{\end{split}}
\newcommand{\bed}{\begin{displaymath}}
\newcommand{\eed}{\end{displaymath}}

\def\kl3g{K_{\ell3\gamma}}
\def\ke3g{K_{e3\gamma}}
\newcommand{\mpi}{M_\pi^2}
\newcommand{\mk}{M_K^2}
\newcommand{\me}{M_\eta^2}
\newcommand{\eq}{\,=\,}
\newcommand{\no}{\nonumber}

\newcommand{\Order}{{\cal O}}
\newcommand{\Lagr}{{\cal L}}
\numberwithin{equation}{section}

\newcommand{\had}{\operatorname{had}}
\newcommand{\elm}{\operatorname{em}}
\newcommand{\SD}{\operatorname{SD}}
\newcommand{\IB}{\operatorname{IB}}
\newcommand{\ubar}{\overline{u}}
\newcommand{\dbar}{\overline{d}}
\newcommand{\sbar}{\overline{s}}

\newcommand{\MeV}{{\,{\rm MeV}}}

\newcommand{\Eg}{E_\gamma^*}
\newcommand{\Ecut}{E_\gamma^{\rm cut}}
\newcommand{\Ep}{E_\pi^*}
\newcommand{\Ee}{{E_e^*}}
\newcommand{\te}{\theta_{e\gamma}^*}
\newcommand{\tep}{\theta_{e\pi}^*}
\newcommand{\tng}{\theta_{\nu\gamma}^*}
\newcommand{\tecut}{\theta_{e\gamma}^{\rm cut}}

\newcommand{\scs}{\, , \,}
\newcommand{\sem}{\, ; \,}
\newcommand{\nnnl}{\nonumber\\}
\newcommand{\fs}{\, . \,}

\newcommand{\vmn}{V_{\mu\nu}}

\begin{document}


\title{Aspects of radiative \boldmath{$K^+_{e3}$} decays}
\titlerunning{Aspects of radiative $K^+_{e3}$ decays}

\author{
Bastian Kubis\inst{1}, Eike H. M\"uller\inst{1,2}, J\"urg Gasser\inst{3}, and Martin Schmid\inst{3} 
}  

\institute{
   Helmholtz-Institut f\"ur Strahlen- und Kernphysik,
   Universit\"at Bonn, Nussallee 14--16, D-53115 Bonn, Germany
\and
   School of Physics, University of Edinburgh,
   King's Buildings, Mayfield Road, Edinburgh EH9 3JZ, United Kingdom
\and
   Institut f\"ur theoretische Physik, Universit\"at Bern, 
   Sidlerstrasse 5, CH-3012 Bern, Switzerland
}

\authorrunning{B. Kubis, E. H. M\"uller, J. Gasser, and M. Schmid}

\date{
}

\abstract{
We re-investigate the radiative charged kaon decay
$K^\pm \to \pi^0 e^\pm \nu_e \gamma$ [$\ke3g^\pm$] in chiral perturbation theory,
merging the chiral expansion with Low's theorem.
We thoroughly analyze the precision of the predicted branching ratio
relative to the non-radiative decay channel.
Structure dependent terms and their impact on differential decay distributions
are investigated in detail, and the possibility to see effects of the 
chiral anomaly in this decay channel is emphasized.
\PACS{
      {13.20.Eb}{Radiative semileptonic decays of $K$ mesons}
      \and
      {11.30.Rd}{Chiral symmetries}
      \and
      {12.39.Fe}{Chiral Lagrangians}
     } 
} 

\maketitle

\section{Introduction}\label{sec:intro}

A particularly useful concept for the investigation of radiative processes
is the decomposition of the transition amplitudes in an inner 
bremsstrahlung (IB) 
and a structure dependent (SD) part. 
The bremsstrahlung part is non-vanishing (in fact, divergent)
in the soft photon limit, and it can be, according to Low's theorem~\cite{low},
expressed entirely in terms of the corresponding non-radiative amplitude
and derivatives thereof; in terms of Feynman diagrams, it corresponds
to photon radiation off the external charged particles.  
Only the structure dependent part contains genuinely new information 
on the photon coupling to intermediate hadronic states.  

Low's theorem was applied to radiative $K_{\ell3}$ decays in Refs.~\cite{FFS70a,FFS70b}.
As is often the case for processes where bremsstrahlung is not forbidden 
by some mechanism, the IB part of the amplitude was found to be largely dominant
in the partial decay widths, and so several of the earlier studies 
(see e.g.\ Ref.~\cite{doncel}) rather concentrated on precision tests 
of soft photon theorems.  
Only with the advent of modern high statistics kaon decay experiments
has it become feasible to measure a relatively rare
decay channel to the required precision such as to even find effects 
of structure dependent contributions.

Experiments typically concern the branching ratio relative to the corresponding
non-radiative process $K^\pm \to \pi^0 e^\pm \nu_e$ [$K_{e3}^\pm$],
\beq
\rR\left(\Ecut,\,\tecut\right) \eq
\frac{\Gamma\left(\ke3g^\pm,\, \Eg > \Ecut,\, \te>\tecut \right)}
     {\Gamma\left(K_{e3}^\pm\right)} ~,  
\label{Rdef}
\eeq
where a minimal photon energy $\Ecut$ as well as a minimal 
photon--positron opening angle $\tecut$
(both in the kaon rest frame) are specified.

\begin{sloppypar}
Experimental results for the relative branching ratios $\rR$ for $\ke3g^\pm$
are shown in Table~\ref{tab:experiments}, where we display the
dependence on the angular cut in degrees, although in some of the original
publications only the cosine is mentioned.
\begin{table}
\centering
\caption{Experimental values of $\rR$ for the decay 
$\ke3g^\pm$.
A single value for the angle cut $\tecut$ denotes
a minimal photon--lepton angle, while ranges refer
to minimal and maximal values. 
\label{tab:experiments}}
\medskip
\renewcommand{\arraystretch}{1.5}
\begin{tabular}{cccccc}
\hline
Ref.           & $\Ecut$  & $\tecut$           & events & $\rR \times 10^2$ \\ 
\hline
\cite{bolotov} & $10\MeV$ & $26^\circ\!-53^\circ$ &   192 & $0.56\pm 0.04$ \\ 
\cite{barmin,PDG}  & $10\MeV$ & $26^\circ\!-53^\circ$ &    82 & $0.46\pm 0.08$ \\
\cite{barmin,PDG}  & $10\MeV$ & $10^\circ$           &    82 & $1.51\pm 0.25$ \\
\cite{istra}   & $10\MeV$ & [none]           &  3852 & $1.69\pm 0.03\pm0.07$ \\  
\cite{istra}   & $10\MeV$ & $26^\circ\!-53^\circ$ &  1423 & $0.48\pm 0.02\pm0.03$ \\  
\cite{istra}   & $30\MeV$ & $20^\circ$           &       & $0.63\pm 0.02\pm0.03$ \\  
\hline 
\end{tabular}
\renewcommand{\arraystretch}{1.0}
\end{table}
Ref.~\cite{bolotov} also includes values for $\rR$ with different photon energy cuts
(but always for the same angular interval).
One should note that the results of the analysis in Ref.~\cite{barmin} 
explicitly refer to the decay probability of an inner bremsstrahlung radiative process.
We have omitted earlier experimental data in Table~\ref{tab:experiments}, 
with statistics determined by less than 20 candidate events, see 
Refs.~\cite{romano,ljung}.
\end{sloppypar}

New experimental efforts that ought to supersede most of the previous results
are under way at NA48/2~\cite{Na48priv} and KEK-E470~\cite{kek},
and ought to record more detailed
information than just the partial widths, in particular precise photon energy
distributions in order to extract information on structure dependent terms.
They are likely to significantly surpass the statistics obtained
in the most recent measurements of the corresponding neutral kaon decay mode
$K_L \to \pi^\mp e^\pm \nu_e \gamma$ [$\ke3g^0$] \cite{Alavi01,Lai04,Alex04},
therefore more precise results on structure dependent terms in $\ke3g^+$ 
ought to be feasible than obtained for those in $\ke3g^0$ 
in the pioneering investigation of Ref.~\cite{Alavi01}. 

\begin{sloppypar}
The appropriate theoretical tool to match such experimental refinements
is chiral perturbation theory (ChPT)~\cite{Weinberg,GLChPT}, the effective
field theory of the Standard Model at low energies.  Radiative $K_{\ell3}$ decays
were calculated up to order $p^4$ in the chiral expansion in Ref.~\cite{BEG}
(see also Ref.~\cite{holstein} for an earlier tree-level calculation),
and in Ref.~\cite{GKPV} this work was extended and combined with the 
useful aspects of the Low expansion for $\ke3g^0$.  
It is the aim of the present article
to apply the ideas and methods developed in Ref.~\cite{GKPV} to $\ke3g^+$
decays.  In Sect.~\ref{sec:form}, we present the necessary formalism
on the $\ke3g^+$ decay amplitudes, in particular the IB--SD separation.
ChPT results on structure dependent terms are discussed analytically and
numerically in Sect.~\ref{sec:chpt}, where in particular 
we present the complete $\Order(p^6)$ results for the axial amplitudes.
In Sect.~\ref{sec:r}, we derive our prediction for the ratio of branching ratios
$\rR$ and give a detailed account of the uncertainties in such a prediction.
Sect.~\ref{sec:distributions} discusses the possibility to extract
information on structure dependent terms from differential decay distributions,
and in particular on the axial anomaly in this decay.
Finally, we summarize our findings in Sect.~\ref{sec:conclusions}.
\end{sloppypar}

In this work, we disregard the interesting topic of $T$-odd correlations
in $\ke3g^+$ decays, which was discussed in detail in Refs.~\cite{Bra02,Bra03,Todd}.

\section{Formalism}\label{sec:form}

In the following, we consider the decay channel
\bea
K^+(p)&\to& \pi^0(p')\,e^+(p_e)\,\nu_e(p_\nu)\,\gamma(q)
\quad\quad [K^+_{e3\gamma}]\label{Kp}
\eea
and its charge conjugate mode.  We only study the emission of a real
photon, that is $q^2=0$.

\subsection{The decay amplitude}

\begin{figure}
\vskip 2mm
  \centering
  \includegraphics[width=7.5cm]{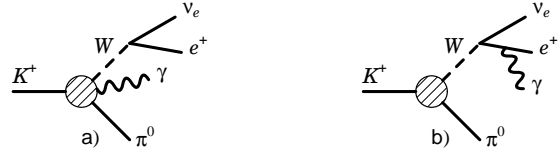}
  \caption{Diagrams describing $\ke3g^+$ decay}
  \label{fig:diag_K+}
\end{figure}

The transition matrix element for $\ke3g^+$ has the form
\bea
T(K_{e3\gamma}^+) &=&
\frac{G_F}{\sqrt{2}}\,e\,V^*_{us}\,\epsilon^\mu(q)^* \notag \\
&\times& \biggl[\bigl(V_{\mu\nu} - A_{\mu\nu}\bigr) \, 
  \bar{u}(p_\nu)\,\gamma^\nu\,(1-\gamma_5)\,v(p_e) \notag \\
&+& \frac{F_\nu}{2 p_e q}\,\bar{u}(p_\nu)\, \gamma^\nu\,(1-\gamma_5)\,
\bigl(m_e-\sh p_e-\sh q\bigr)\,\gamma_\mu\,v(p_e)\biggr] \notag \\[2mm]
&\doteq& \epsilon^\mu(q)^* \,M_\mu ~. \label{me_Kpg}
\eea
The relevant diagrams are displayed in  Fig.~\ref{fig:diag_K+}.
The first term of \eqref{me_Kpg} corresponds to diagram a), which
includes bremsstrahlung off the charged kaon, 
while the second one corresponds to the
radiation off the positron, represented by diagram b). 
We have introduced the hadronic tensors $V_{\mu\nu}$ and $A_{\mu\nu}$,
\bea
I_{\mu\nu} &\eq& i\,\int d^4 x\,e^{iqx}\,
\langle \pi^0(p')|T\,V^{\rm em}_\mu(x)\,I^{\rm had}_\nu(0)|K^+(p)\rangle ~\sem\no\\
\quad I &\eq& V,A  ~,
\label{tensor}
\eea
with
\begin{eqnarray}
V_\nu^{\had} &=& \sbar \gamma_\nu u ~,\quad A_\nu^{\had} = \sbar \gamma_\nu \gamma_5 u ~,\notag\\
V_\mu^{\elm} &=& (2\ubar\gamma_\mu u-\dbar\gamma_\mu d-\sbar\gamma_\mu s)/3 ~,
\end{eqnarray}
whereas $F_\nu$ is the $K^+_{e3}$ matrix element
\beq
F_\nu\eq  \langle \pi^0(p')|V^{\rm had}_\nu(0)|K^+(p)\rangle ~. 
\label{me_Kp}   
\eeq
The tensors $V_{\mu\nu}$ and $A_{\mu\nu}$ satisfy the Ward identities
\beq
q^\mu V_{\mu\nu} \eq F_\nu ~,\qquad
q^\mu A_{\mu\nu} \eq 0 ~,
\label{WI}
\eeq
which imply gauge invariance of the total amplitude
\eqref{me_Kpg},
\beq
q^\mu M_\mu\eq 0 ~.
\label{WI_me}
\eeq  

\begin{sloppypar}
The $\ke3g^+$ amplitude can be decomposed into 
inner bremsstrahlung and structure dependent parts.
We require these two amplitudes to be separately gauge invariant,
and the SD part to be of order $q$ and higher in an expansion
in powers of the photon momenta.
According to Low's theorem~\cite{low}, the IB terms, which comprise
the part of the amplitude non-vanishing for small photon momenta
(and in particular the infrared divergent pieces),
are given entirely in 
terms of the $K^+_{e3}$ form factors $f_+$, $f_1$ defined by
\begin{eqnarray}
  F_\nu(t) &=& \frac{1}{\sqrt{2}}\Big(2p'_\nu f_+(t)+(p-p')_\nu f_1(t)\Big) ~,\label{eq:decf}
\end{eqnarray}
with $t=(p-p')^2$.
[We use the form factors $f_+$, $f_1$ instead of the more conventional
$f_+$, $f_-=f_1-f_+$ in order to be able to write the IB amplitude
in a slightly more compact form below.]
This splitting of the matrix element implies a corresponding splitting of the 
hadronic tensors $V_{\mu\nu}$ and $A_{\mu\nu}$.
The axial correlator $A_{\mu\nu}$ consists of structure dependent parts only,
and can be expressed in terms of four scalar functions $A_i$, $i=1\ldots4$,
\begin{eqnarray}
  A_{\mu\nu} 
  &=& \frac{i}{\sqrt{2}}\biggl[
  \epsilon_{\mu\nu\rho\sigma}\bigl(A_1 \,p'^\rho q^\sigma + A_2 \,q^\rho W^\sigma\bigr) \label{deca}\\
  && +\, \epsilon_{\mu\lambda\rho\sigma} \, p'^\lambda q^\rho W^\sigma
  \biggl( \frac{A_3}{M_K^2-W^2}\,W_\nu+A_4\,p'_\nu\biggr)
  \biggr] ~,\notag
\end{eqnarray}
where $W = p-p'-q$. 
(We use the convention $\epsilon_{0123} = +1$.)
Note that in comparison to Refs.~\cite{BEG,GKPV}, we have factored
out the kaon pole explicitly in the definition of the structure function $A_3$. 

The decomposition of the vector correlator reads
\beq\label{eq:decompIBSD}
  V_{\mu\nu} = V_{\mu\nu}^{\IB} + V_{\mu\nu}^{\SD} ~,
\eeq
where the SD piece is chosen such that
\begin{xalignat}{2}
  q^\mu V_{\mu\nu}^{\IB} &= F_\nu(t)~, & \quad q^\mu V_{\mu\nu}^{\SD} &= 0 ~.
\end{xalignat}
For a given choice of  $V_{\mu\nu}^{\rm SD}$, the structure dependent
part of the decay amplitude $T$ in \eqref{me_Kpg} is defined to be
\beq \begin{split}
T^{\rm SD} &\eq \frac{G_F}{\sqrt{2}}\,e\,V^*_{us}\,\epsilon^\mu(q)^*
\left(V_{\mu\nu}^{\rm SD}-A_{\mu\nu}\right) \times \\
& \hskip 2.5cm 
\times \bar{u}(p_\nu)\,\gamma^\nu\,(1-\gamma_5)\,v(p_e) ~,
\end{split}
\eeq
whereas the bremsstrahlung part is $T^{\rm IB}=T-T^{\rm SD}$.
\end{sloppypar}

It remains to explicitly construct the decomposition \eqref{eq:decompIBSD}.
This is done explicitly in Appendix~\ref{app:WI}, where we derive the form of $V_{\mu\nu}^{\IB}$ 
in terms of $f_+$ and $f_1$ as
\begin{eqnarray}
  V_{\mu\nu}^{\IB} &=& \frac{1}{\sqrt{2}}
  \bigg[
  \frac{p_\mu}{pq}\bigl(2p'_\nu f_+(W^2) + W_\nu f_1(W^2)\bigr)\notag\\ &&
  +\;\;\frac{W_\mu}{qW}\bigl(2p'_\nu \triangle f_++W_\nu \triangle f_1\bigr)
  +g_{\mu\nu}f_1(t)
  \bigg]\notag ~,\\
  \triangle f_i &=& f_i(t)-f_i(W^2) ~,\quad i = +,1 ~. \label{eq:VIB}
\end{eqnarray}

A slightly different representation for the IB amplitude was already derived in
Refs.~\cite{FFS70a,FFS70b}.  It differs from the one given above
by terms of order $q$.  An important feature of the form \eqref{eq:VIB} is
the fact that it contains all singularities at $pq=0$ in the sense
that the complete residue at $pq=0$, which is a non-trivial function 
of the momenta $p$, $p'$, and $q$, is contained in it,
and no terms like $(qW)^2/pq$ (which is formally of order $q$)
are shifted to the structure dependent tensor.

The structure dependent part of the vector correlator can also be expressed 
in terms of four scalar functions $V_i$, $i=1\ldots4$, 
in a basis of gauge invariant tensors according to~\cite{poblaguev}
\begin{eqnarray}
  V_{\mu\nu}^{\SD} &=& \frac{1}{\sqrt{2}}\Bigl[
  V_1 \,(p'_\mu q_\nu-p'q\,g_{\mu\nu}) +
  V_2 \,( W_\mu q_\nu- qW\,g_{\mu\nu}) \notag\\
  &&\qquad + V_3 \,\bigl(qW \, p'_\mu W_\nu  - p'q \, W_\mu W_\nu \bigr) \notag\\
  &&\qquad + V_4 \,\bigl(qW \, p'_\mu p'_\nu - p'q \, W_\mu p'_\nu\bigr)
  \Bigr] ~. \label{eq:ViSD}
\end{eqnarray} 

\subsection{Kinematics}

We briefly collect the essential information on the kinematics of this decay.
The Lorentz invariant amplitudes $A_i$, $V_i$ 
are functions of three scalar variables, which we often take to be
\beq
s=(p'+q)^2 ~,~~ t=(p-p')^2 ~,~~ u=(p-q)^2 ~.
\eeq
These variables are particularly useful in the discussion of the analytic properties of $V_i$, $A_i$.
The  physical region  in $\ke3g$ decays can be represented as follows:
for fixed $W^2$, the variables $s$, $t$, and $u$ vary in the range
\bea\label{eq:physicalstu}
W^2&\leq &t\leq (M_K-M_\pi)^2  ~, \nnnl 
s_-&\leq &s \leq s_+ ~, \nnnl
s_\pm&=&M_\pi^2-\frac{1}{2t}\bigl(t+M_\pi^2-M_K^2\bigr)(t-W^2)\nnnl
&&\pm\frac{1}{2t}
\lambda^{1/2}(t,M_K^2,M_\pi^2)\lambda^{1/2}(t,0,W^2) ~,\nnnl
s+t+u&=&M_K^2+M_\pi^2+W^2 ~,
\eea
where $\lambda$ is the usual K{\"a}ll{\'e}n function, 
\bea
\lambda(x,y,z)&=&x^2+y^2+z^2-2(xy+xz+yz)\fs
\eea
Note that $M_\pi^2 = M_{\pi^0}^2$ denotes the neutral pion mass in this work.
Varying the invariant mass squared $W^2$ 
of the lepton pair in the interval
\bea
m_e^2 \leq W^2 \leq (M_K-M_\pi)^2
\eea
generates the  region covered by $s$, $t$, $u$ in $\ke3g$ decays. 
Instead of $s$, $t$, $u$, we also use
\beq\label{eq:varEEW}
p q/M_K = \Eg ~,~~ p p'/M_K=\Ep ~,~~ W^2 = 
(p_e+p_\nu)^2 ~,
\eeq
where $\Eg$, $\Ep$ are the photon and the pion energy 
in the kaon rest frame. These variables are often useful when 
discussing partial decay widths. 

\begin{sloppypar}
For the four-body decay $\ke3g^+$, one needs five independent variables to describe 
the kinematics of the decay completely.  We choose the two additional scalar products
\beq
p p_e/M_K = \Ee ~,~~ x = p_e q/M_K^2 ~,
\label{var2}
\eeq
where $\Ee$ is the positron energy in the kaon rest frame.
The dimensionless variable $x$ is related to the angle
$\te$ between the photon and the positron:
\beq
x M_K^2 \eq \Eg\,\Bigl(\Ee - \sqrt{\Ee^2-m_e^2}\,\cos{\te}\Bigr) ~.
\label{theta}
\eeq
The smallness of the electron mass leads to a near-vanishing of $x$
for collinear electron and photon momenta, and hence to a near-singularity,
which is avoided by cutting on the angle $\te$.

The total decay rate is given by
\begin{align}
&\Gamma(K^+ \to \pi^0 e^+ \nu \gamma) \eq \label{rate}\\
& \hskip 1cm
\frac{1}{2 M_K (2\pi)^8}\,\int
d_{\rm LIPS}(p;\,p',p_e,p_\nu,q) \, \sum_{\rm spins} \left|T\right|^2 ~, \no
\end{align}
where $d_{\rm LIPS}(p;p',p_e,p_\nu,q)$ is the Lorentz invariant 
phase space element for the $\ke3g^+$ process.\footnote{For the decay 
of a particle of momentum $p$ into $n$
  particles of momenta $p_1,\dots,p_n$, one has
\[
d_{\rm LIPS}(p;p_1,\dots,p_n) \eq
 \delta^4\Bigl(p-\sum_{i=1}^n p_i\Bigr)\prod_{k=1}^n\frac{d^3 p_k}{2p_k^0}~.
\]}
When performing the traces over the spins, we work with 
massless spinors, such that the form factors
$A_3$, $V_3$, and $f_1$ drop out, and 
$\sum_{\rm spins} \left|T\right|^2$ is a bilinear form 
of the invariant amplitudes 
$V_i$, $A_i$, $i=1,\,2,\,4$, and $f_+$. 
The explicit result is displayed in Appendix~\ref{app:traces}.
\end{sloppypar}

\section{Structure dependent terms in ChPT}\label{sec:chpt}

\subsection{Analytical results at order \boldmath{$p^4$}}

In Ref.~\cite{BEG}, the chiral expansion was carried out to order $p^4$
for both the neutral and the charged decay modes (one-loop order). 
We do not describe that calculation in any detail and 
only quote the result, adjusted to the separation between bremsstrahlung and
structure dependent terms.
The axial amplitudes are given in terms of the Wess--Zumino--Witten anomaly term~\cite{WZ,Witten}, 
the result is
\beq
  A_1 = -4 A_2 = A_3 = -\frac{1}{2\pi^2 F^2} ~, ~~
  A_4 = 0  \quad
  \bigl[\Order(p^4)\bigr] ~. \label{axial}
\eeq
The vector structure functions $V_i$ were shown to be given as
\beq \begin{split}
V_1 &= \sqrt{2}I_2                         ~, \quad
V_2 = -\frac{\sqrt{2}}{qW}\bigl(I_1+p'q \, I_2\bigr)  ~,   \\
V_3 &= \frac{\sqrt{2}}{qW}\bigl(I_3-f_2^+(W^2)\bigr) ~, \quad
V_4 = 0 \qquad \bigl[\Order(p^4)\bigr] ~,
\end{split} \label{chiralVi}
\eeq
where the functions $I_i$, $f_2^+(W^2)$ are defined in Ref.~\cite{BEG}.

It was pointed out in Ref.~\cite{BEG} that these functions are real throughout the physical region.
In fact, cuts only start at $t=(M_K+M_\pi)^2$, $W^2=(M_K+M_\pi)^2$, therefore
far from the phase space boundaries, which is why the $V_i$ at this order
are smooth and well-behaved, and even constant to a high degree of accuracy.
They can therefore be approximated by a seemingly drastic simplification, 
namely by their values at the special kinematical point $s=M_\pi^2$, 
$u=M_K^2$, $t=W^2=0$. This corresponds to an expansion to leading order
in the photon momentum $q$, plus setting $t=0$.
The rather compact and simple result reads
\begin{align}
V_1 &= -\frac{8}{F^2}\bar{L}_9 - \frac{(1-x)^{-2}}{32\pi^2F^2} \biggl\{ 
  \frac{1}{3}\left(53-25x+2x^2\right) 
\no\\ & \hskip 0.45cm
+ \left(1+x-x^2+x^3\right)\frac{\log x}{2(1-x)} 
\no\\ & \hskip 0.45cm
- \left(127-93x+21x^2-x^3\right)\frac{\log y}{2(1-x)} \biggr\}
+\Order(q,t) ~, \no\\ 
V_2 &= -\frac{4}{F^2} \left( \bar{L}_9 +\bar{L}_{10} \right)
\no\\ & \hskip 0.45cm 
- \frac{(1+x)(1-x)^{-2}}{64\pi^2F^2} \biggl\{ 
  1+x + \frac{2x \log x}{1-x}  \biggr\} 
\no\\ & \hskip 0.45cm 
- \frac{(1-x)^{-3}}{32\pi^2F^2} \biggl\{ 
  \frac{166}{3}(9-4x)+(77-x)\frac{x^2}{3}
\no\\ & \hskip 0.45cm
+ x(3+2x) \frac{\log x}{1-x} 
- 9(12-x)(4-x)^2 \frac{\log y}{1-x}  \biggr\} 
\no\\ & \hskip 0.45cm
+\Order(q,t) ~, 
\no\\
V_3 &= - \frac{(1-x)^{-4}}{32\pi^2F^2\mk} \biggl\{ 
  \frac{2611}{3}-13x(34-5x)-\frac{4}{3}x^3 
\no\\ & \hskip 0.45cm
+ x(2+3x+x^2) \frac{\log x}{1-x} 
- 27(7-x)(4-x)^2 \frac{\log y}{1-x}  \biggr\} 
\no\\ & \hskip 0.45cm
+\Order(q,t) ~,\label{Vit=0}
\end{align}
where $x=M_\pi^2/M_K^2$, $y=M_\eta^2/M_K^2$, 
and we have made frequent use of the Gell-Mann--Okubo relation.
See \eqref{barLi} for a definition of the scale-independent low-energy 
constants $\bar L_9$, $\bar L_{10}$. 
We remark that the expressions for $V_1$ and $V_3$ in \eqref{Vit=0}
are identical to the equivalent approximations in the neutral kaon
decay mode quoted in Ref.~\cite{GKPV}.

\subsection{Analytical results at order \boldmath{$p^6$}}\label{sec:p6}

\begin{sloppypar}
As the above-described results for the structure functions $V_i$, $A_i$
at $\Order(p^4)$ are the leading contributions in the chiral expansion,
and as the chiral expansion involving strangeness does not necessarily 
converge very fast, 
it is mandatory to understand the structure of higher-order corrections
in order to give realistic estimates of the structure dependent terms.
\end{sloppypar}

The analytic structure of the tensors $V_{\mu\nu}$, $A_{\mu\nu}$ was investigated
in detail in Refs.~\cite{GKPV,Todd}.  The main results of those investigations are:
\begin{enumerate}
\item 
Due to strangeness conservation, cuts in the variables $t$, $u$, $W^2$ 
can all start only at $(M_K+M_\pi)^2$, which is far outside the physical region.
As in the case of the vector structure functions at $\Order(p^4)$, 
these cuts are expected to hardly affect the momentum dependence of the $V_i$, $A_i$
inside the decay region, where they can therefore be approximated by polynomials.
\item \begin{sloppypar}
At higher order in the chiral expansion, there are two-- and three--pion cuts
in the $s$-channel that make the structure functions complex.  These cuts 
were discussed in great detail in Ref.~\cite{Todd}.  
Due to the photon in the final state, all the diagrams developing imaginary parts
have a $P$-wave characteristic, i.e.\ the imaginary parts rise only slowly 
above threshold, and a cusp-like structure in the real parts is smoothed out
($\propto (s-4M_\pi^2)^{3/2}$ in the case of the two--pion cuts).
For our purposes, the real parts of the structure functions can therefore
still be regarded as ``smooth''. \end{sloppypar}
\end{enumerate}
The main purpose of the analysis of higher-order corrections is therefore 
to see how large the corrections to the averaged (constant) structure functions
might be.

\subsubsection{Complete order $p^6$ corrections to the axial amplitudes}\label{sec:p6axial}

We have calculated the complete $\Order(p^6)$ corrections to
the axial structure functions $A_1$, $A_2$, and $A_4$. 
The contribution of $A_3$ to the squared matrix element
is always suppressed by a factor of $m_e^2/M_K^2 \approx 10^{-6}$ 
and is therefore neglected.
The generic structure is as follows:
\begin{align}
A_1 &\eq -\frac{1}{2\pi^2F_\pi F_K} \Bigl\{1+
 S_1(s) + T_1(t) + U_1(u) +X_1  \Bigr\}~, \notag \\
A_2 &\eq \frac{1}{8\pi^2F_\pi F_K} \Bigl\{ 1+ 
  S_2(s)+T_2(t) + U_2(u) +X_2 \Bigr\} ~, \notag \\
A_4 &\eq -\frac{C_{4A}}{F_\pi F_K}~.\label{Aistruct}
\end{align}
The explicit forms for the various loop functions as well as the
combinations of low-energy constants entering the expressions
\eqref{Aistruct} can be found in
Appendix~\ref{app:axial}. 
In particular, the constant terms $X_{1,2}$ in \eqref{Aistruct}
contain terms proportional to the 
strange quark mass, and are therefore potentially big.
We remark that by renormalizing the order $p^4$ amplitudes
according to $F^2 \to F_\pi F_K$, all dependence on the
low-energy constants $L_4$ and $L_5$ is absorbed 
in the physical meson decay constants.

\subsubsection{Polynomial corrections to the vector amplitudes}\label{sec:p6vector}

A complete evaluation of the vector structure functions at order $p^6$
requires a full two-loop calculation and is beyond the scope of this article.
In order to gain a basic idea about potential higher-order corrections,
we content ourselves with an investigation of just polynomial contributions.
Note that the loop contributions with cuts in the physical region 
are known to be tiny at this order~\cite{Todd}.

The polynomial part for the $V_i$ can be calculated from the Lagrangian
$\Lagr_{6}$~\cite{L6Scherer,L6Bij}.     
For the form factors $V_1$, $V_2$, it contains all possible terms linear in
$s$, $t$, $u$, $W^2$, $\mk$, $\mpi$. 
The numerically potentially largest corrections for $V_1$, $V_2$ turn out to
be the terms suppressed by a factor $M_K^2/(4\pi F)^2$ with respect to the
leading $\bar{L}_9$, $\bar{L}_{10}$ contributions.   
The (leading) polynomial contributions for $V_3$, $V_4$, 
which only appear at $\Order(p^6)$, are constant. 

We have furthermore calculated the contributions of the form $L_i \times L_j$
at order $p^6$.  In analogy to what was done above for the axial structure functions,
we find that a renormalization of the order $p^4$ couplings according to 
$F^2 \to F_\pi F_K$ takes care of all such terms.  
This information is used for the numerical evaluation of the structure functions 
below, where such a normalization is chosen in order to minimize higher-order corrections.

\subsection{Numerical evaluation of structure dependent terms}\label{sec:sdnum}

\begin{table}
\centering
\caption{
  Values for the structure dependent terms as given by
  $\Order(p^4)$ and estimated from $\Order(p^6)$ ChPT.  
  The symbols $\av{V_i}$, $\av{A_i}$ denote the averages of the {real parts}
  of the $V_i$, $A_i$ over phase space, in units of $M_K$.
  For the $V_i$, we also show the approximation given in \eqref{Vit=0}.
  For the error bands, see discussions in main text.\label{tab:viai}}
\medskip
\renewcommand{\arraystretch}{1.4}
\begin{tabular}{crclcrcl}
\hline
  & \mcc{$\Order(p^4)$} & \eqref{Vit=0} & \mcc{$\Order(p^6)$} \\
\hline
$\av{V_1}$ & $-1.24$\nl$\pm$\nl$0.004$&$-1.23$& $-1.24$\nl$\pm$\nl$0.4$ \\ 
$\av{V_2}$ & $-0.19$\nl$\pm$\nl$0.007$&$-0.21$& $-0.19$\nl$\pm$\nl$0.2$ \\
$\av{V_3}$ & $-0.02$\nl$\pm$\nl$0.001$&$-0.02$&        \nl$\pm$\nl$0.1$ \\
$\av{V_4}$ & $ 0   $\nl$   $\nl$     $&$ 0   $&        \nl$\pm$\nl$0.1$ \\
\hline
$\av{A_1}$ & $-1.19$\nl$   $\nl$     $&       & $-1.29$\nl$\pm$\nl$0.4$ \\ 
$\av{A_2}$ & $ 0.30$\nl$   $\nl$     $&       & $ 0.33$\nl$\pm$\nl$0.1$ \\
$\av{A_3}$ & $-1.19$\nl$   $\nl$     $&       &        \nl$   $\nl$   $ \\
$\av{A_4}$ & $ 0   $\nl$   $\nl$     $&       &        \nl$\pm$\nl$0.3$ \\
\hline 
\end{tabular}
\renewcommand{\arraystretch}{1.0}
\end{table}
In Table~\ref{tab:viai}, we summarize the numerical evaluation of the various
structure dependent terms.  We use the parameters and numerical values
specified in Appendix~\ref{app:numerical}.  In particular, we neglect
any variation in the order $p^4$ low-energy constants $L_9$, $L_{10}$,
the effect of which should be generously covered by the uncertainty in the order $p^6$ 
contributions estimated below.

The averages $\av{V_i}$, $\av{A_i}$ in the first column of Table~\ref{tab:viai}, 
referring to the values of the structure dependent terms at order $p^4$, 
are obtained by integrating the structure functions over phase space and
dividing by the phase space volume; the quoted uncertainties are the
$1\sigma$ errors of this averaging procedure.
These are therefore no realistic estimates of the uncertainties of the mean values,
but only serve as an illustration to what extent the approximation 
of the $V_i$ being constant at this order is justified.  
It is seen that the variation within physical phase space is absolutely 
negligible.  This is, on the one hand, due to the absence of cuts,
but also to the fact that $V_1$ and $V_2$ are (at the scale of the $\rho$ mass)
numerically dominated by counterterm contributions, which are in turn 
necessarily constant at this order.
For comparison, we also show the values of the $V_i$ at the special kinematical
point $s=M_\pi^2$, $u=M_K^2$, $t=W^2=0$ as given in \eqref{Vit=0}, 
which are consistent with this picture.

A numerical assessment of the corrections at next-to-leading order ($p^6$)
is much more difficult, essentially due to the large number of unknown 
low-energy constants.  
For the axial structure functions that we have calculated completely,
we proceed as follows:  by averaging the (real parts of)
the loop contributions at a scale $\mu=M_\rho$ over phase space, 
we calculate the shifted mean values $\av{A_i}$; the counterterm contributions
are added as the essential uncertainty.  
This uncertainty is determined by examining the scale dependence of the 
combined counterterm contributions.
If the scale is varied such that the corresponding logarithms change by one,
we find for the counterterm parts of $A_{1,2}$
\bea
A_{1,\rm ct} &\eq& \mp\frac{1}{192\pi^4F_\pi^2 F_K^2} 
\Bigl\{ 14 \bigl(M_K^2 - M_\pi^2\bigr) + 4 s + t + u 
\Bigr\}~, \notag\\
A_{2,\rm ct} &\eq& \pm\frac{1}{768\pi^4F_\pi^2 F_K^2} 
\Bigl\{ 17 M_K^2 - 9 M_\pi^2 
  - 7 t + 4 u \Bigr\} ~.~~ \label{eq:Aiscale}
\eea
As there are no loop contributions to $A_4$, the corresponding
counterterm combination is separately scale independent, 
and we have to utilize an even simpler order-of-magnitude estimate
with the result
\beq
A_{4,\rm ct} \eq \pm \frac{16}{(4\pi)^4F_\pi^2F_K^2} ~. \label{eq:A4scale}
\eeq
In order to obtain simple error ranges for  $\av{A_{1,2}}$, 
we again average the momentum dependent terms over phase space. 
It is seen that the largest contributions to the uncertainty stem 
from constant terms $\propto M_K^2$.
In this sense, within the accuracy at which these structure functions can 
presently be predicted,
we can even neglect the momentum dependence and approximate the structure 
functions  by constants.  

The results thus obtained for the $A_i$ are displayed in the third column of
Table~\ref{tab:viai}.
We note that the uncertainties for $A_{1,2}$ are of the size of typical
chiral SU(3) corrections of about 30\%.

\begin{sloppypar}
For the $V_i$, we use the same order-of-magnitude arguments concerning
higher-order contributions as in Ref.~\cite{GKPV}.  The uncertainty
for the dominant structure function $V_1$ is estimated to be of the order
of 30\%; as $V_2$ is suppressed at leading order, we scale its uncertainty
by a factor of 2.  The (constant) counterterm contributions to $V_{3,4}$
are estimated by dimensional arguments similar to that for $A_4$ discussed above.
All the numbers are collected in the third column of Table~\ref{tab:viai}.
\end{sloppypar}

\section[The ratio           $\rR$ ]
        {The ratio \boldmath{$\rR$}}\label{sec:r}

The ratio $\rR$ defined in \eqref{Rdef} is a particular useful quantity to
consider in $\ke3g$ decays, as it is both the quantity that is naturally
measured in experiment (as opposed to the branching ratio $\Gamma(\ke3g)/\Gamma_{\rm all}$),
and can be predicted in a clean way theoretically.

We repeat here the corresponding discussion of $\rR$ for the neutral kaon decay
in Ref.~\cite{GKPV} for the decay channel $K^+_{e3\gamma}$.  As the layout of the
formalism is analogous to the neutral channel, we shall be rather
brief and only comment in more detail on the numerical results.

For the moment, we neglect radiative corrections and isospin breaking and denote
$\rR$ in the absence of virtual and real photon corrections by $\rr$,
\beq
\alpha^{-1} \rr = \bigl[ \alpha^{-1} \rR \bigr]_{\alpha=0} ~.
\eeq
We will comment on radiative corrections in Sect.~\ref{sec:radcorr}.

\begin{sloppypar}
We define the quantity $\ol{\rm SM}$ by
\beq 
\Gamma(K^+_{e3\gamma}) 
\doteq  
  \frac{4\alpha M_K^5 G_F^2 |V_{us}|^2}{(2\pi)^7} f_+(0)^2
   \int d_{\rm LIPS} \; \ol{\rm SM} ~,
\eeq
such that $\int d_{\rm LIPS} \ol{\rm SM}$ is dimensionless and contains no further
(electroweak) coupling constants.  By factoring out $f_+(0)^2$, $\ol{\rm SM}$ 
only contains the normalized form factor $\bar f_+(t) = f_+(t)/f_+(0)$. 
The non-radiative width $\Gamma(K^+_{e3})$ is conventionally written as
\beq \begin{split}
\Gamma(K^+_{e3}) &= \int dy \, dz \,\rho(y,z) ~,\\
\rho(y,z) &= 
\frac{M_K^5 G_F^2 |V_{us}|^2}{256\pi^3} f_+(0)^2 A(y,z) \bar{f}_+(t)^2 ~,
\end{split}\eeq
where $y=2pp_e/M_K^2$, $z=2pp'/M_K^2$, and 
\beq \begin{split}
A(y,z)&\eq 4(z+y-1)(1-y)+r_e(4y+3z-3) \\ & \hskip 0.45cm
-4r_\pi+r_e(r_\pi-r_e) ~,
\end{split} \eeq
with $r_e=m_e^2/M_K^2$, $r_\pi=M_\pi^2/M_K^2$.
The expression for
$\rr$ in terms of these reduced phase space integrals
is then of the same form as for the neutral decay mode~\cite{GKPV},
\beq
\rr = \frac{8\alpha}{\pi^4} \frac{\int d_{\rm LIPS} \, \ol{\rm SM}}
                               {\int dy\,dz\,A(y,z)\bar{f}_+(t)^2} ~. \label{r}
\eeq
We will occasionally also refer to a ratio $\rr^{\rm IB}$, which is understood
to be calculated according to \eqref{r}, with all structure dependent 
contributions in $\ol{\rm SM}$ omitted.
\end{sloppypar}

\subsection{Phase space integrals}

\begin{sloppypar}
Assuming\footnote{Note that, while we use $M_\pi = M_{\pi^0}$ elsewhere in this text,
the expansion of the form factor $f_+$ is conventionally normalized to the 
charged pion mass $M_{\pi^\pm}$.}
\beq
\bar{f}_+(t) = 1+\lambda_+ \frac{t}{M_{\pi^\pm}^2}+\lambda''_+ \frac{t^2}{M_{\pi^\pm}^4} ~,
\eeq
one may expand the integral in the denominator according to 
\begin{align} 
I &\eq \int dy\,dz\,A(y,z)\bar{f}_+(t)^2 \label{kl3density} \\
&\eq a_0 + a_1 \lambda_+ + a_2 \left(\lambda_+^2+2\lambda''_+\right)
+ a_3 \lambda_+ \lambda''_+ + a_4 {\lambda''_+}^2 ~, \no
\end{align}
where the numerical values for the $a_i$ are collected in
Table~\ref{tab:ai}.  
\begin{table}
\centering
\caption{Coefficients for the $K^+_{e3}$ phase space integral.\label{tab:ai}}
\medskip
\renewcommand{\arraystretch}{1.4}
\begin{tabular}{ccccc}
\hline
$a_0$ & $a_1$ & $a_2$ & $a_3$ & $a_4$ \\
\hline
0.09653 & 0.3337 & 0.4618 & 3.189 & 6.278 \\
\hline
\end{tabular}
\renewcommand{\arraystretch}{1.0}
\end{table}
We point out that we have used
the physical (charged) kaon and (neutral) pion masses everywhere, 
such that the kinematics in the phase space integral correspond to the physical situation.
\end{sloppypar}

Similarly we can expand the phase space integral for the radiative decay in
terms of $K_{e3}$ form factor parameters according to
\begin{align} 
I^\gamma &\eq \int d_{\rm LIPS} \, \ol{\rm SM} \label{kl3gdensity} \\
&\eq b_0 + b_1 \lambda_+ + b_2 \lambda_+^2+ b_3 \lambda''_+
+ b_4 \lambda_+ \lambda''_+ + b_5 {\lambda''_+}^2 ~. \no
\end{align}
The integral $I^\gamma$ and the coefficients $b_i$ depend on the 
experimental cuts $\Ecut$, $\tecut$.  
\begin{table}
\centering
\caption{Coefficients for the $K^+_{e3\gamma}$ phase space integral.\label{tab:bip}}
\medskip
\renewcommand{\arraystretch}{1.4}
\begin{tabular}{cccccc}
\hline
$b_0^{\rm IB}$ &  $b_1^{\rm IB}$ & $b_2$ & $b^{\rm IB}_3$ &  $b_4$ & $b_5$ \\
\hline
$ 1.019$ & $ 3.98$ & 
$ 5.81 $ & $11.83$ &
$41.9  $ & $84.8$\\
\hline
\mc{$b_0^{\rm SD}$} & \mc{$b_1^{\rm SD}$} & \mc{$b^{\rm SD}_3$}  \\
\hline
\mc{$-0.012\pm0.004$} & \mc{$-0.03\pm0.01$} & \mc{$ -0.10\pm0.03$}  \\
\hline
\end{tabular}
\renewcommand{\arraystretch}{1.0}
\end{table}
The numerical results for the standard cuts $\Ecut=30\MeV$, $\tecut=20^\circ$ 
are displayed in Table~\ref{tab:bip}. 
Where applicable, the coefficients have been decomposed into their
bremsstrahlung and their structure dependent parts.  The relative size of the
structure dependent contributions as predicted by ChPT is very similar to the
neutral kaon decay channel, they reduce the width by about 1\%. 
The uncertainties quoted in Table~\ref{tab:bip} 
refer to the estimated higher-order contributions in the structure dependent terms
as discussed in Sect.~\ref{sec:sdnum}.

\subsection{Form factor dependence of \boldmath{$\rr$}}

We can now study the dependence of $\rr$ ($\rr^{\rm IB}$) on the form factor parameters
$\bar \lambda_+ = \lambda_+/\lambda_+^c$, $\bar \lambda''_+ = \lambda''_+/(\lambda_+^c)^2$,
where we choose $\lambda_+^c = 0.0294$ as a central value for the slope parameter 
$\lambda_+$. 
We expand $\rr$ according to 
\begin{align}
\rr\left(\bar{\lambda}_+,\bar{\lambda}''_+\right)
\eq \rr(1,0) \Bigl\{ 1
&+ c_1 \, \left(\bar{\lambda}_+ -1 \right) 
+ c_2 \, \left(\bar{\lambda}_+ -1 \right)^2
\no\\ & 
+ c_3 \, \bar{\lambda}''_+
 + \ldots \Bigr\} ~, \label{Nr}
\end{align}
and $\rr^{\rm IB}$ accordingly (with expansion coefficients $c_i^{\rm IB}$).
We show the numbers for the coefficients $c_i$, $c_i^{\rm IB}$ in Table~\ref{tab:ci}.
We find that although there is a significant cancellation between
the $\lambda_+$, $\lambda''_+$ dependence of numerator and denominator, the
cancellation is not quite as complete as for the $K^0_{e3(\gamma)}$ channel~\cite{GKPV}.
Going from a point-like form factor ($\lambda_+=0$) to the physical one
$\lambda_+=\lambda_+^c$ induces a change of about 1.3\% in $\rr$ or $\rr^{\rm IB}$.
\begin{table}
\centering
\caption{
  Coefficients for the $\bar{\lambda}_+$, $\bar{\lambda}''_+$ 
  dependence of $\rr^{\rm IB}$, $\rr$.  
  The error margins for the $\rr$ coefficients are due to 
  uncertainties in higher-order contributions to the structure dependent terms. \label{tab:ci}}
\medskip
\renewcommand{\arraystretch}{1.4}
\begin{tabular}{cccc}
\hline
$\rr^{\rm IB}(1,0)\cdot 10^2$ & $c_1^{\rm IB}\cdot 10^3$ 
& $c_2^{\rm IB}\cdot 10^4$ & $c_3^{\rm IB}\cdot 10^4$  \\
\hline
$0.640$ & $12.0$ & $-5.4$ & $16.6$ \\
\hline
$\rr(1,0)\cdot10^2$ & $c_1\cdot10^3$ & $c_2\cdot 10^4$ & $c_3\cdot 10^4$  \\
\hline
$0.633\pm0.002$ & $12.5\pm0.4$ & $-5.4\pm0.3$ & $16.9\pm0.4$ \\
\hline
\end{tabular}
\renewcommand{\arraystretch}{1.0}
\end{table}

On the other hand, $\rr$ and $\rr^{\rm IB}$ are remarkably stable 
within the range of uncertainties of the measured slope parameters.  
Even if the latest experimental results that determine $\lambda''_+$
\cite{ISTRAff,KTeVff,NA48ff,KLOEff} do not agree perfectly with each other, all these
results combined show a strong (anti)\-cor\-rela\-tion between $\lambda_+$ and $\lambda''_+$
(see e.g.\ Fig.~8 in Ref.~\cite{KLOEff}).
If we vary $(\lambda_+,\lambda''_+)$ in the parameter space indicated by the different $1\sigma$
ellipses of Refs.~\cite{ISTRAff,KTeVff,NA48ff,KLOEff}, we find that $\rr$ differs from
$\rr(1,0)$ by less than a permille.  
With the present knowledge of the form factor $f_+(t)$, it is therefore already possible
to predict $\rr$ to excellent precision, and the biggest uncertainty (of the order of 0.4\%)
stems from unknown higher-order corrections in the structure-dependent terms.

\subsection{Dependence on the experimental cuts}

We briefly study the dependence of the parameters in \eqref{Nr} on the
experimental cuts by displaying their values for the alternative cuts
$\Ecut=10\MeV$, $\tecut=10^\circ$  in
Table~\ref{tab:ci_cuts}.
\begin{table*}
\centering
\caption{$\rr^{\rm IB}$, $\rr$ for different values of the experimental cuts on $\Ecut$, $\tecut$,
  as well as coefficients for the $\bar{\lambda}_+$, $\bar{\lambda}''_+$ dependence of $\rr$.
  The error margins for the $\rr$ coefficients are due to 
  uncertainties in higher-order contributions to the structure dependent terms.
  \label{tab:ci_cuts}}
\medskip
\renewcommand{\arraystretch}{1.4}
\begin{tabular}{ccccccc}
\hline
$\Ecut$ & $\tecut$ &
$\rr^{\rm IB} \cdot 10^2$ & 
$\rr \cdot 10^2$ & 
$c_1 \cdot 10^3$ & 
$c_2 \cdot 10^4$ & 
$c_3 \cdot 10^4$ \\
\hline 
30\,MeV & $20^\circ$         & 0.640 
                            & $0.633\pm 0.002$ & $12.5\pm 0.4$ & $-5.4\pm 0.3$ & $16.9\pm 0.4$ \\
30\,MeV & $10^\circ$         & 0.925 
                            & $0.918\pm 0.002$ & $11.1\pm 0.3$ & $-4.7\pm 0.2$ & $15.0\pm 0.3$ \\
10\,MeV & $20^\circ$         & 1.211 
                            & $1.204\pm 0.002$ & $7.5\pm 0.2$ & $-3.2\pm 0.2$ &  $10.1\pm 0.2$ \\
10\,MeV & $10^\circ$         & 1.792 
                            & $1.785\pm 0.002$ & $6.7\pm 0.2$ & $-2.8\pm 0.1$ & $9.0\pm 0.1$ \\
10\,MeV & $26^\circ-53^\circ$ & 0.554 
                            & $0.553\pm 0.001$ & $5.7\pm 0.1$ & $-2.4\pm 0.1$ & $7.5\pm 0.1$ \\
\hline 
\end{tabular}
\renewcommand{\arraystretch}{1.0}
\end{table*}
In addition, for historical reasons we also show results for the
angle range $26^\circ \leq \te \leq 53^\circ$ in combination with
$\Ecut = 10\MeV$, compare Table~\ref{tab:experiments}.
$\rr$, $\rr^{\rm IB}$ of course depend strongly on the cut values.
The dependence on the form factor parameters, however, remains very mild
also for these different cuts.
In particular, the ratio $\rr$ is even more stable with respect to 
variations of the form factor parameters, i.e.\ the variation of $\lambda_+$,
$\lambda''_+$ in the range of the latest experimental results,
as described above, leads to a change in $\rr$ below the permille level.
As the difference between the coefficients $c_i$ and $c_i^{\rm IB}$ 
is as small for all cuts as for the standard ones discussed in the previous
subsection, we refrain from showing the  $c_i^{\rm IB}$ in Table~\ref{tab:ci_cuts}.

\subsection{Isospin breaking and radiative corrections}\label{sec:radcorr}

As we can predict the ratio $\rr$ to surprising precision of about half a percent, 
using Low's theorem, the experimentally available information on the $K_{e3}$ 
form factor $f_+$, and ChPT for the structure dependent terms, we have to comment
on isospin breaking corrections, generated by real and virtual photons and
the light quark mass difference $m_u-m_d\neq 0$, which may clearly have an impact
at the percent level.

As radiative corrections comprise the radiation of (additional) soft real photons,
we have to specify the precise meaning of the ratio $\rR$ as defined in \eqref{Rdef},
in the presence of virtual and real photons.  We define the denominator to be 
the fully inclusive width $K^+ \to \pi^0 e^+ \nu_e (n\gamma)$, 
where $(n\gamma)$ denotes any number of photons of arbitrary energy.
In analogy, the numerator in \eqref{Rdef} refers to the measurement
of radiative $K^+_{e3}$ decays with at least one photon fulfilling the cut
requirements $\Eg > \Ecut$, $\te > \tecut$, plus arbitrary additional photons.

A complete calculation of radiative corrections in $\rR$ is beyond the scope of this
article.  Below, we argue that the most sizeable effects can easily be taken care of,
and that the unknown corrections can reasonably be expected to be small.

\begin{enumerate}
\item 
As was already argued in Ref.~\cite{GKPV}, the large short distance
enhancement factor $S_{EW} \propto \log M_Z/M_\rho$ that renormalized the Fermi coupling constant 
$G_F$~\cite{marcianosirlin,kl3radA,kl3radB} applies identically in numerator and denominator
of $\rR$, so it cancels in the ratio.

\item 
\begin{sloppypar}
In Ref.~\cite{kl3radA}, isospin breaking corrections 
in $K_{e3}$ were calculated in chiral perturbation theory up-to-and-including
$\Order\bigl(p^2e^2,p^2(m_u-m_d)\bigr)$. 
Parts of these can be collected in a modified value for $f_+(0)$, which
still cancels in $\rR$.
In addition, the parameters $a_i$ describing the slope parameter expansion
of the $K_{e3}$ phase space integral are shifted, the modified 
coefficients of the expanded phase space integral are given in
Table~\ref{tab:aiem}. 
As in Ref.~\cite{kl3radA}, the contribution of (single) real photon radiation
was restricted to pion/electron momenta in agreement with non-radiative kinematics,
we re-modify the coefficient $a_0$ to be fully inclusive again, increasing it by
$0.51\%$.
\end{sloppypar}
\begin{table}
\centering
\caption{Coefficients for the $K^+_{e3}$ phase space integral,
including corrections of $\Order(\alpha, m_u-m_d)$.
The values for $a_{1,2}$ are taken from Ref.~\cite{kl3radA}.
We are grateful to V.~Cirigliano for providing us with the numbers for $a_{3,4}$,
which are not included in that reference. For $a_0$, see text.
\label{tab:aiem}}
\medskip
\renewcommand{\arraystretch}{1.4}
\begin{tabular}{ccccc}
\hline
$a_0$ & $a_1$ & $a_2$ & $a_3$ & $a_4$ \\
\hline
0.09583 & 0.3287 & 0.4535 & 3.124 & 6.136 \\
\hline
\end{tabular}
\renewcommand{\arraystretch}{1.0}
\end{table}

\item
The part of this estimate that is necessarily incomplete concerns
isospin breaking corrections to the numerator of $\rR$. 
Radiative corrections that are potentially large are those enhanced
by electron mass singularities.  As we have defined the numerator of $\rR$
to be \emph{inclusive} with respect to additional photon radiation,
such electron mass singularities should, according to the KLN theorem~\cite{Kinoshita,LN}, 
be taken care of by evaluating the running coupling constant at the scale
of the kaon mass,
$\alpha \to \alpha\bigl(1+\frac{\alpha}{3\pi}\log(M_K^2/m_e^2)\bigr)$.
In Ref.~\cite{kl3radA}, a part of these large logarithms are absorbed into $f_+(0)$,
which reduces the correction in the numerator to 
$\alpha \to \alpha\bigl(1+\frac{\alpha}{12\pi}\log(M_K^2/m_e^2)\bigr)$.

\item
We expect the remaining, non-enhanced, radiative corrections to be small, of the order of
$\alpha/\pi$.  As a conservative estimate, we assign a relative uncertainty due 
to these of the size $\pm \Delta_{\rm em} = \pm 5\alpha/\pi \approx \pm 0.01$. 

\item
Finally, corrections in the numerator due to the light quark mass difference
that are not local and absorbed in $f_+(0)$ are expected to be tiny and are neglected.

\end{enumerate}
Summing everything up, we find that the corrections in the denominator of $\rR$
plus the remaining electron mass logarithms enhance $\rR$ by 1\%.  
The estimated uncertainty in other radiative corrections is larger than that
from higher-order chiral corrections in the structure dependent terms.  
Our combined predictions, for the various cut combinations discussed before,
are summarized in Table~\ref{tab:Rfinal}.
\begin{table}
\centering
\caption{Combined results for $\rR$ for various cut combinations. \label{tab:Rfinal}}
\medskip
\renewcommand{\arraystretch}{1.4}
\begin{tabular}{ccc}
\hline
$\Ecut$ & $\tecut$ & $\rR\cdot 10^2$ \\
\hline
30\,MeV & $20^\circ$         & $0.640 \pm 0.008$ \\
30\,MeV & $10^\circ$         & $0.928 \pm 0.011$ \\
10\,MeV & $20^\circ$         & $1.217 \pm 0.014$ \\
10\,MeV & $10^\circ$         & $1.804 \pm 0.021$ \\
10\,MeV & $26^\circ-53^\circ$ & $0.559 \pm 0.006$ \\
\hline
\end{tabular}
\renewcommand{\arraystretch}{1.0}
\end{table}

We briefly compare the predictions in Table~\ref{tab:Rfinal} to the experimental
results shown in Table~\ref{tab:experiments}.  We find very good agreement
with Ref.~\cite{bolotov} as well as with the standard cuts result in 
Ref.~\cite{istra}, while the values for $\rR$ reported in Ref.~\cite{barmin,PDG}
as well as the one for the angular cut range $26^\circ-53^\circ$ in Ref.~\cite{istra}
are a bit below the theoretical values, though disagreement hardly exceeds $1\sigma$
deviation.  
Clearly, forthcoming even more precise determinations of $\rR$~\cite{Na48priv,kek} 
are very welcome.

In Ref.~\cite{istra}, the ratio $\rR$ is also given with no cut on $\te$, see
Table~\ref{tab:experiments}.  It is difficult to provide a precise theoretical
value of $\rR$ for this situation, because the electron mass singularity
renders the numerical integration over phase space unstable.
While our calculation suggests a rather bigger value for $\rR$ than what is
quoted  in Table~\ref{tab:experiments}, we refrain from working out a precise result.
We do not, therefore, provide a final number in Table~\ref{tab:Rfinal}.

\section{Structure dependent terms\\ in differential rates}\label{sec:distributions}

\subsection{Photon energy distribution}\label{sec:Egdist}

It is obvious from the results of the previous section that
a precise measurement of the ratio $\rR$
is insufficient to determine structure dependent terms in $\ke3g^+$: 
the shift in $\rR$ is tiny and easily overwhelmed by the uncertainty
in radiative corrections.  It seems therefore more promising to use
the more detailed experimental information contained in differential
distributions for this purpose.  
In a pioneering study, the KTeV collaboration has attempted such
an extraction of structure dependent contributions for $\ke3g^0$
decays~\cite{Alavi01} by measuring the photon energy distribution, 
and their results were analyzed in detail
in Ref.~\cite{GKPV}.  Here we provide the theoretical basis for 
such an analysis for $\ke3g^+$, together with the chiral prediction
for the outcome of such an extraction.

Among the various differential rates that may potentially be
investigated, the one with respect to the photon energy $\Eg$ 
is predestinated for a separation of bremsstrahlung and structure dependent
terms, as the different behavior of both at small $\Eg$ 
is the defining property to distinguish between the two contributions.

In the decomposition of the photon energy spectrum, we neglect
terms that are proportional to structure dependent terms squared, 
and we use the working assumption discussed in Sect.~\ref{sec:chpt} that 
the structure functions can reasonably well be approximated by constants.
The photon energy distribution thus reads
\beq \begin{split}
\frac{d\Gamma}{d\Eg} = \frac{d\Gamma_{\rm IB}}{d\Eg}
&+ \sum_{i=1}^4 \left( \av{V_i} \,\frac{d\Gamma_{V_i}}{d\Eg}
                    + \av{A_i} \,\frac{d\Gamma_{A_i}}{d\Eg} \right) \\
&+\Order\Bigl(|T^{\rm SD}|^2,\,\Delta V_i,\,\Delta A_i\Bigr) \fs
\end{split} \label{specdecomp}
\eeq
Here $d\Gamma_{V_i}/d\Eg$ is defined to be the part of the spectrum
proportional to $V_i$, etc., and $\Delta V_i$, $\Delta A_i$ stand 
for the errors made by the approximation of constant structure functions.

\begin{figure}
\vskip 2mm
  \centering
  \includegraphics[width=0.95\linewidth]{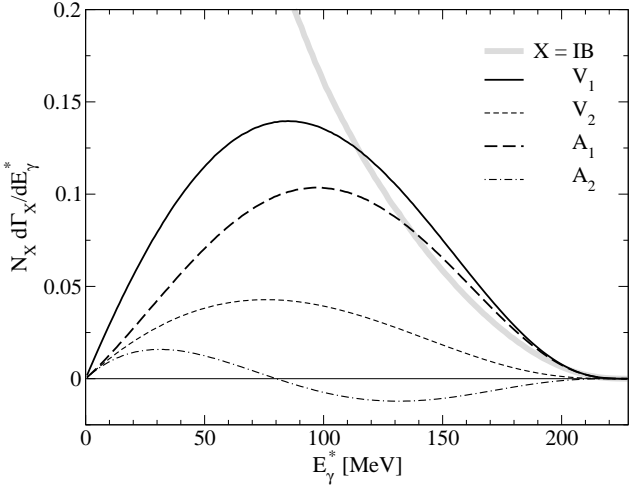}
\vskip 2mm
  \caption{Photon energy distributions from inner bremsstrahlung
  as well as the various structure dependent terms.  The notation
  $d\Gamma_X/d\Eg$ for the various $X$ refers to
  \eqref{specdecomp}.  The normalization factors are
  $N_{V_i,\,A_i}=100\,N_{\rm IB}=10^3M_K/\Gamma(K_{e3})$. 
  We only cut on the electron--photon angle,
  $\tecut=20^\circ$. \label{fig:Egdist}}
\end{figure}
\begin{sloppypar}
We remind the reader that $V_3$, $A_3$ are suppressed by $m_e^2/M_K^2$ and are
therefore essentially unobservable.  Furthermore, we find that
the distributions $d\Gamma_{V_4}/d\Eg$, $d\Gamma_{A_4}/d\Eg$ are considerably
smaller than the other structure dependent parts of the spectrum;
adding to that the observation that $\av{V_4}$, $\av{A_4}$ are suppressed
by two orders in the chiral expansion with respect to $\av{V_{1,2}}$, $\av{A_{1,2}}$,
we neglect these structures and only discuss the effects of $V_{1,2}$, $A_{1,2}$
for simplicity reasons.
\end{sloppypar}

The remaining parts of the photon spectrum $d\Gamma_X/d\Eg$ are shown in
Fig.~\ref{fig:Egdist}.  Note that in order to put all spectra in one plot,
the bremsstrahlung spectrum has been scaled down by two orders of magnitude.
While $d\Gamma_{\rm IB}/d\Eg$ displays the expected $1/\Eg$ singularity
for small photon energies, the structure dependent parts 
$d\Gamma_X/d\Eg$ with $X=V_1,\,V_2,\,A_1$ perturb the pure bremsstrahlung spectrum
all essentially in the same way:  they rise linearly with $\Eg$ for small energies,
are bent down by phase space at maximum photon energies, and display a maximum
between 80~MeV and 100~MeV.  
Even though the strength of the perturbation varies, the shape is nearly the same for all three.
Only the distribution $d\Gamma_{A_2}/d\Eg$ has an additional node in between
and is therefore suppressed.  

\begin{sloppypar}
Similarly to the neutral kaon decay channel~\cite{GKPV}, we find that the
bremsstrahlung spectrum is perturbed by a function $f(\Eg)$, where
\beq\label{eq:fegamma}
f(\Eg)
\;\doteq\;            \frac{d\Gamma_{V_1}}{d\Eg} 
\;\approx\; 3.3 \times\frac{d\Gamma_{V_2}}{d\Eg} 
\;\approx\; 1.3 \times\frac{d\Gamma_{A_1}}{d\Eg}  
~,
\eeq
The information on the SD terms is contained 
in the effective strength  $\av{X}$ that 
multiplies $f(\Eg)$,
\beq \begin{split}
\frac{d\Gamma}{d\Eg} &\,\approx\,
\frac{d\Gamma_{\rm IB}}{d\Eg}  + \av{X} \;f(\Eg)  ~, \\
\av{X} &\eq     \av{V_1} 
        + 0.3 \,\av{V_2}
        + 0.7 \,\av{A_1} \label{Vicombine} ~.
\end{split} \eeq
Taking into account the chiral predictions for the various averaged
structure functions, see Table~\ref{tab:viai}, we find that $\av{X}$
is dominated by $\av{V_1}$ and $\av{A_1}$, both of which contribute
with comparable strength.  Corrections from $\av{V_2}$, $\av{A_2}$ are
suppressed effectively by roughly a factor of 20.
Numerically, we predict the effective strength $\av{X}$ to be
\bea\label{eq:Xnum}
 \av{X} \eq
\left\{ \begin{array}{ll}
 -2.2         & ~\Order(p^4) \\[2mm]
 -2.2 \pm 0.7 & ~\Order(p^6) ~.
\end{array} \right.
\eea
We therefore conclude that a significant part of the structure dependent
photon spectrum is due to the chiral anomaly ($A_1$).  
This is different from the decay $\ke3g^0$~\cite{GKPV}, where $A_1$
vanishes at $\Order(p^4)$, and the contribution of $A_2$ to the photon 
spectrum is comparably suppressed as in Fig.~\ref{fig:Egdist}.
The importance of the chiral anomaly for the existence 
of various contributions to $T$-odd correlations in $\ke3g^+$
was emphasized in Ref.~\cite{Todd}, therefore it plays a significant
role in the phenomenology of this process.
\end{sloppypar}

\subsection{Other distributions}\label{sec:otherdist}

It is particularly interesting to trace back effects of the chiral anomaly even more clearly,
i.e.\ we would like to separate the contributions of $V_1$ and $A_1$
that can only be measured as a (weighted) sum in the photon energy distribution.
This turns out to be rather difficult:  in
many differential distributions, both are practically
indistinguishable from each other and/or from the dominant bremsstrahlung spectrum.
\begin{figure}
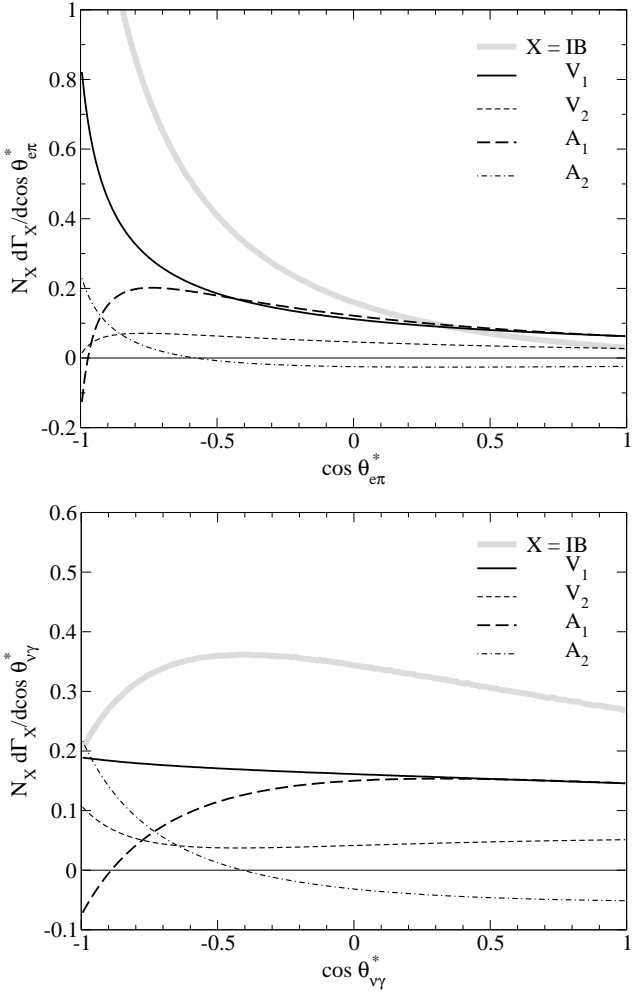

\vskip 2mm
\centering 
\includegraphics[width=0.95\linewidth]{dCepi_bw.eps} 
\vskip 2mm
\includegraphics[width=0.95\linewidth]{dCnug_bw.eps} 
\caption{Distributions with respect to $\cos\tep$, $\cos\tng$ from
  inner bremsstrahlung
  as well as the various structure dependent terms.  The notation
  $d\Gamma_X/d\cos\tep$, $d\Gamma_X/d\cos\tng$,  is chosen in
  analogy to \eqref{specdecomp}.  The normalization factors are
  $N_{V_i,\,A_i}=100\,N_{\rm IB}=10^4/\Gamma(K_{e3})$. 
  Both spectra are shown for the standard cuts $\Ecut=30$~MeV, $\tecut=20^\circ$.
  \label{fig:CTepinugdist}}  
\end{figure}
Two notable exceptions are shown in Fig.~\ref{fig:CTepinugdist}:
the distributions with respect to the (cosines of) the angles
between electron and pion, as well as between neutrino and photon,
show strongly different behavior in backward directions for the parts
of the spectra proportional to $\av{V_1}$ and $\av{A_1}$.  
In particular, $d\Gamma_{A_1}/d\cos\tep$, $d\Gamma_{A_1}/d\cos\tng$
go through zero here.

A strategy for an advanced study of structure dependent terms
that tries to disentangle the two most important SD contributions 
$\av{V_1}$ and $\av{A_1}$ on purely experimental grounds, without
input from ChPT, might therefore
proceed along the following lines:  
\begin{enumerate}
\item determine the effective strength $\av{X} \approx \av{V_1} + 0.7 \av{A_1}$
  from an analysis of the photon energy spectrum; 
\item use this constraint for an additional fit to the angular spectra
  $d\Gamma/d\cos\tep$ and/or $d\Gamma/d\cos\tng$ in order to determine
  $\av{V_1}$ and $\av{A_1}$ separately.
\end{enumerate}

\begin{sloppypar}
The (numerically) subleading structure functions $\av{V_2}$, $\av{A_2}$ are, due to their smallness, 
even more difficult to determine.  In principle, one finds
a similar picture for these as in the case of $K^0_{e3\gamma}$ \cite{GKPV}:
the distribution $d\Gamma_{V_2}/d\Ep$ peaks at lower pion energies
than bremsstrahlung and the dominant structure dependent terms, 
and $d\Gamma_{A_2}/d\cos\te$ produces a slightly enhanced
variation in backward directions.  
However, these effects are expected to be much harder to measure
even than the structure dependent modification of the photon energy spectrum.

\begin{figure}
\vskip 2mm
\centering 
\includegraphics[width=0.95\linewidth]{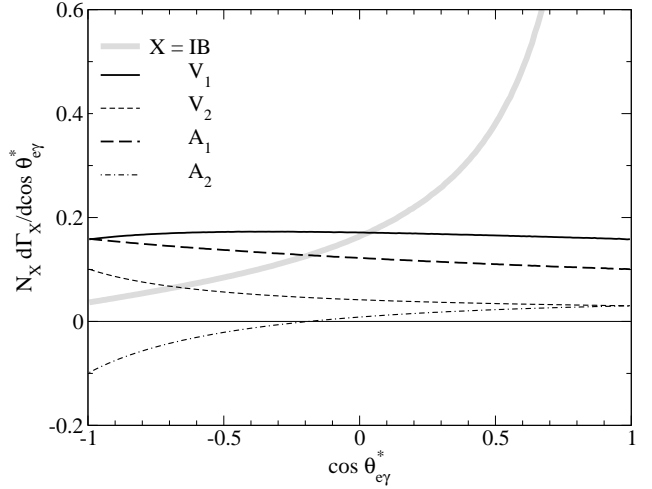} 
\vskip 2mm
\caption{Distributions with respect to $\cos\te$ from
  inner bremsstrahlung
  as well as the various structure dependent terms.  The notation
  $d\Gamma_X/d\cos\te$ is chosen in
  analogy to \eqref{specdecomp}.  The normalization factors are
  $N_{V_i,\,A_i}=100\,N_{\rm IB}=10^4/\Gamma(K_{e3})$. 
  The photon energy cut $\Ecut=30$~MeV was applied.
  \label{fig:CTegdist}}  
\end{figure}

We want to briefly comment on the differential distributions $d\Gamma/d\cos\te$,
the different contributions to which are displayed in Fig.~\ref{fig:CTegdist}.
[Note again that the bremsstrahlung contribution is scaled down by two orders
of magnitude with respect to the structure dependent ones.]
In Ref.~\cite{istra}, a discrepancy between data and Monte Carlo was
found for $\cos\te \approx -1$, and the authors comment that this ``could
possibly be interpreted as a direct emission effect''.  
It is obvious from Fig.~\ref{fig:CTegdist} that ChPT makes this claim untenable:
the effects of structure dependent terms are too small and in particular
far too flat in their angular distributions to produce a visible
peak as seen in Fig.~8 of Ref.~\cite{istra}. 
The resulting total curve IB+SD would still lie within the width of the grey line for the 
IB contribution shown in Fig.~\ref{fig:CTegdist}.
\end{sloppypar}

\section{Conclusions}\label{sec:conclusions}

In this article, we have analyzed various aspects of $\ke3g^+$ decays,
spinning forth previous work on $\ke3g^0$.  Our findings can be summarized as follows:
\begin{enumerate}
\item \begin{sloppypar}
We have constructed a decomposition of the $\ke3g^+$ decay amplitude 
into an inner bremsstrahlung and a structure dependent part, in the absence
of radiative corrections, that guarantees that the SD part has 
no non-trivial analytic properties except for unitarity cuts.  
\end{sloppypar}
\item
By applying this decomposition to the chiral representation of the $\kl3g^+$
amplitude of order $p^4$~\cite{BEG}, we derive the leading chiral predictions
for the structure functions.
The axial structure functions that are given in terms of the Wess--Zumino--Witten
anomaly are constant, while the vector structure functions, although given
in terms of complicated loop functions, are shown to be
free of cuts in the physical region and therefore smooth and very close to constant.
\item
We consider higher-order corrections of order $p^6$ to the structure functions,
in particular, we present the complete order $p^6$ expressions for the axial terms.
Although cuts in the physical region due to intermediate two--pion states appear,
their cusp behavior is shown to be of $P$-wave type and therefore suppressed,
such that the real parts of the $A_i$ are still smooth to a reasonable approximation.
The size of (incomplete) higher-order polynomial corrections
to the vector structure functions is estimated to be of natural size 
for chiral SU(3) corrections.
\item 
We have investigated the stability of the prediction of the relative branching ratio 
$\rR$, normalized to the non-radiative width, with respect to $K_{e3}$ form factor
parameters.  Within the range of uncertainty of the available experimental information
on these parameters, $\rR$ is shown to be remarkably stable at the permille level.
Structure dependent terms reduce $\rR$ by about 1\%. 
Our combined prediction for experimental cuts of $\Ecut = 30\MeV$, $\tecut=20^\circ$
is
\beq
\rR = (0.640 \pm 0.008)\times 10^{-2} ~,
\eeq
where the uncertainty is dominated by unknown radiative corrections.
Predictions for other cut values were also given in Table~\ref{tab:ci_cuts}.
\item
We have discussed how to extract a linear combination of structure dependent
terms from an experimental analysis of the differential rate $d\Gamma/d\Eg$.
In contrast to $\ke3g^0$, the axial anomaly contributes significantly
to the perturbation of the bremsstrahlung spectrum, such that effects may 
actually be visible in $\ke3g^+$.  
As a second step, we have pointed out that angular distributions
of the type $d\Gamma/d\cos\tep$ and/or $d\Gamma/d\cos\tng$ may allow for a disentanglement
of the two dominant structure dependent terms $V_1$ and $A_1$.
\end{enumerate}
It would be extremely interesting and rewarding to see the various
predictions tested by the modern high-statistics kaon decay experiments
such as NA48/2~\cite{Na48priv} or KEK-E470~\cite{kek},
in particular to find unambiguous
signals for the physics behind the structure dependent terms.

\begin{acknowledgement}
\textit{Acknowledgements.} 

\begin{sloppypar}  
We thank Nello Paver and Michela Verbeni for collaboration
in an early stage of this project.
We are grateful to Stefano Venditti for useful communications. 
This work was supported in parts 
by the EU Integrated Infrastructure Initiative Hadron Physics Project 
   (contract number RII3-CT-2004-506078),
by DFG (SFB/TR 16, ``Subnuclear Structure of Matter''), 
by the Swiss National Science Foundation, by RTN, BBW-Contract No. 01.0357,
   and EC-Contract HPRN--CT2002--00311 (EURIDICE).
J.~G. is grateful to the Alexander von Humboldt-Stiftung and to
the Helmholtz-Gemeinschaft for a grant,
and he thanks the HISKP, where part of this work was performed,
 for the warm hospitality.
E.~H.~M.\ thanks the ``Studienstiftung des deutschen Volkes'' 
for supporting his studies.
\end{sloppypar}  
\end{acknowledgement}

\renewcommand{\thefigure}{\thesection.\arabic{figure}}
\renewcommand{\thetable}{\thesection.\arabic{table}}

\begin{appendix}

\setcounter{figure}{0}
\setcounter{table}{0}
\section[Inner bremsstrahlung in $\ke3g^+$ decays]
        {Inner bremsstrahlung in \boldmath{$\ke3g^+$} decays}\label{app:WI}

In this appendix we discuss the separation of the hadronic tensor
$V_{\mu\nu}$ into an IB and a SD part. 
\begin{figure}[hb]
  \centering
  \includegraphics[width=7.5cm]{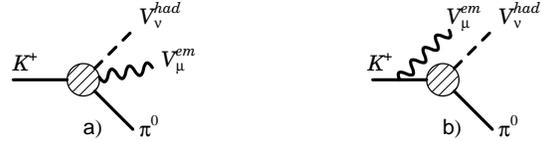}
  \caption{Diagrams for $\vmn$, evaluated in the framework of ChPT. The
    hatched blobs denote one-particle irreducible graphs.}
  \label{fig:IB}
\end{figure}
The relevant diagrams can be grouped in two classes,  displayed in
Fig.~\ref{fig:IB}. The hatched blobs denote one-particle irreducible
contributions. The diagram b) generates a pole in 
the variable $u=(p-q)^2$,  at $u=M^2_K$, corresponding to the
intermediate $K^+$ state.
We isolate the contribution of this pole by writing 
\beq
V_{\mu\nu} \eq \tilde{V}_{\mu\nu} + \frac{1}{\sqrt{2}}\frac{p_\mu}{p q}\,
\left[2p'_\nu f_+(W^2) + W_\nu f_1(W^2) \right] ~,
\label{pole}
\eeq
where $W=p-p'-q$. In the following, we assume that this is the only
singular part at $q=0$ in the tensor $V_{\mu\nu}$, or, in other words, that
$\tilde V_{\mu\nu}$ is regular at $q=0$. This is the only assumption in the
derivation of the final expression for the IB term. We have checked that it
is true at one-loop order in ChPT, and we  see no reason why it should
not be correct to any order, and thus true in QCD.

We write this regular part as
\beq\begin{split}
\tilde V_{\mu\nu} \eq \frac{1}{\sqrt{2}} &\bigl[
v_0\,g_{\mu\nu}+v_1\, p'_\mu q_\nu+
v_2\,W_\mu q_\nu  +v_3\, p'_\mu W_\nu  \\
&+ v_4\, p'_\mu p'_\nu  + v_5\, W_\mu p'_\nu
 + v_6\, W_\mu W_\nu  \bigr] ~. 
\end{split}\eeq
The Ward identity \eqref{WI} generates three
conditions on $\tilde V_{\mu\nu}$ that can be written as
\beq \begin{split} 
v_0 + v_1\, p'q + v_2\, qW &\eq f_1 ~,\\
      v_3\, p'q + v_6\, qW &\eq \triangle f_1 ~,\\
      v_4\, p'q + v_5\, qW &\eq 2\triangle f_+  ~,
\end{split} \eeq
with
\beq
\triangle f_i \eq f_i(t)-f_i(W^2) ~,\quad i = +,1 ~.
\eeq
The first equation can be solved for $v_0$. Furthermore, we set
\beq
v_5 \eq \frac{2\triangle f_+}{qW}+\tilde v_5 ~,~
v_6 \eq \frac{\triangle f_1}{qW}+\tilde v_6 ~.
\eeq
and find
\beq \begin{split}
v_4\, p'q+\tilde v_5\, qW \eq 0 ~,\\
v_3\, p'q +\tilde v_6\,qW \eq 0 ~.
\end{split} \label{eq:twoequations} \eeq
We use the fact that the Lorentz invariant amplitudes $v_i$ are
defined for any value of the kinematic variables $p'q$, $qW$, and that the
amplitudes are assumed to be non-singular at $p'q=0$.  It then follows
that $\tilde v_{5,6}$  are proportional to $p'q$,
\beq
\tilde v_{5,6} \eq -p'q \,\tilde v_{4,3} ~,
\eeq
where the sign and the numbering is  chosen for convenience. Finally, we obtain
\beq
v_{3,4} \eq qW \,\tilde v_{3,4} ~.
\eeq
Collecting the results, we find that $V_{\mu\nu}^{\rm SD}$ can be written in the form
displayed in \eqref{eq:ViSD}, with
\beq
(V_1,V_2,V_3,V_4) \eq (v_1,v_2,\tilde v_3,\tilde v_4) ~.
\eeq
\eqref{Vit=0} contains the explicit expression of
the form factors $V_i$ in the limit $q=0$, $t=0$, illustrating 
that they indeed are non-singular at $q=0$ at next-to-leading
order in ChPT, as mentioned above.

\setcounter{figure}{0}
\setcounter{table}{0}
\section{Traces}\label{app:traces}

Here, we give the explicit expression for the sum over spins in
$|T|^2$ in the limit where
the relevant traces are evaluated at $m_e=0$.
We write
\beq \begin{split}
N^{-1}\sum_{\rm spins}|T|^2 
&\eq
a_1 \,f_+(t)^2+a_2\,f_+(t) \delta f_+ +a_3\,\delta f_+ ^{\, 2}     
\\[-2mm] & \,+\,
\sum_{i=1}^4\Bigl[\bigl(b_i  \,\text{Re} V_i
                       +b^5_i\,\text{Re} A_i\bigr)f_+(t) 
\\ & \hskip 1cm +
                  \bigl(c_i  \,\text{Re} V_i
                       +c^5_i\,\text{Re} A_i\bigr)\delta f_+\Bigr] 
\\[3mm] & \,+\,
\Order\bigl({\rm Im}{V_i},{\rm Im}{A_i},V_i^2,A_i^2,V_i A_i\bigr) ~,
\end{split}\label{eq:traces} 
\eeq
with
\beq \begin{split}
\delta f_+ &\eq M_K^2(q W)^{-1}\left[f_+(t)-f_+(W^2)\right] ~, \\[2mm]
N &= {8 \pi\alpha G_F^2|V_{us}|^2M_K^2} ~.
\end{split} \eeq
For the neglected terms proportional to imaginary parts
in the structure functions $V_i$, $A_i$, see Ref.~\cite{Todd}.
With this convention for $N$, 
the right hand
side in \eqref{eq:traces} is dimensionless. 
In the limit $m_e=0$, we immediately have
\beq
b_3\eq b_3^5\eq c_3\eq c_3^5\eq 0~.
\eeq
We use the abbreviations
\beq
\begin{array}{lllll}
z\, p   p'   =a \scs&  
z\, p   q    =b \scs&
z\, p   p_e  =c \scs&  
z\, p   p_\nu=d \scs\\[2mm]  
z\, p'  q    =e \scs&  
z\, p'  p_e  =f \scs&  
z\, p'  p_\nu=g \scs&
z\, p_e q    =h \scs\\[2mm] 
z\, p_\nu q  =j \scs&  
z\, p_e p_\nu=k \scs& 
z\, p   W    =l \scs&  
z\, p'  W    =m \scs\\[2mm]
z\, q   W    =n \scs& 
z\, M_\pi^2=r   \scs& 
z\,=M_K^{-2}    \scs&
\end{array}
\eeq
and decompose all the coefficients according to 
$a_i = \hat a_i \, \bar a_i$ etc., where the prefactors 
$\hat a_i\,,\;\hat b_i\,\ldots$ are collected in
Table~\ref{tab:factors}. 
\renewcommand{\arraystretch}{1.4}
\begin{table}
\centering
\caption{Prefactors that  multiply the $\bar a_i\,,\;\bar b_i$ etc.
  \label{tab:factors}
}
\medskip
\begin{tabular}{cccccc}
\hline
$\hat a_1$ & $4/(b^2h)$ &
$\hat b_1$ & $4/(bhz)$  & $\hat b^5_1$ & $4/(bhz)$ \\
$\hat a_2$ & $4/(b^2h)$ &
$\hat b_2$ & $4/(bhz)$  & $\hat b^5_2$ & $4/(bhz)$ \\
$\hat a_3$ & $4/b^2$    &
$\hat b_4$ & $2/(bhz^2)$& $\hat b^5_4$ & $2/(hz^2)$    \\
\hline
& & 
$\hat c_1$ & $4/(bz)$   & $\hat c^5_1$ & $4/(bz)$    \\
& &
$\hat c_2$ & $4/(bz)$   & $\hat c^5_2$ & $4n/(bz)$   \\
& &
$\hat c_4$ & $4/(bz^2)$ & $\hat c^5_4$ & $1/z^2$      \\
\hline
\end{tabular}
\renewcommand{\arraystretch}{1.0}
\end{table}
We obtain the following expressions for the coefficients $\bar a_i$,
$\bar b_i$ and so on:
\bea
\bar a_1 &=&  - 2\,a\,b\,g\,h + 2\,b^2 f\,g + 2\,b^2 g\,e 
- b^2 j\,r 
\nnnl &&
- b^2 k\,r + 4\,b\,c\,f\,g + 2\,b\,c\,g\,e 
- b\,c\,j\,r 
\nnnl &&
- 2\,b\,c\,k\,r + b\,d\,h\,r 
- 2\,f\,g\,h + h\,k\,r  \scs
\nnnl[0.75mm]
\bar a_2 &=& 2\,a\,b\,g\,h\,n + 4\,b^2 f\,g\,k - 2\,b^2 g\,h\,m 
+ 2\,b^2 g\,k\,e 
\nnnl&&
+ b^2 h\,k\,r - b^2 j\,k\,r
- 2\,b^2 k^2 r - 4\,b\,c\,f\,g\,n 
\nnnl&&
- 2\,b\,c\,g\,n\,e + b\,c\,j\,n\,r 
+ 2\,b\,c\,k\,n\,r - b\,d\,h\,n\,r 
\nnnl&&
- 4\,b\,f\,g\,h\,l 
+ 2\,b\,h\,k\,l\,r 
+ 4\,f\,g\,h\,n - 2\,h\,k\,n\,r
  \scs
\nnnl[0.75mm]
\bar a_3 &=&  - 4\,b^2 f\,g\,k + 2\,b^2 k^2 r 
+ 4\,b\,f\,g\,l\,n 
\nnnl &&
- 2\,b\,k\,l\,n\,r - 2\,f\,g\,n^2  + k\,n^2 r  \scs
\nnnl[0.75mm]
\bar b_1 &=&  - a\,f\,h\,j - a\,g\,h^2  + b\,f^2 j + b\,f\,g\,h 
- b\,f\,k\,e 
\nnnl &&
+ 2\,b\,g\,h\,e - b\,h\,j\,r + c\,g\,h\,e + d\,f\,h\,e  \scs
\nnnl[0.75mm]
\bar b_2 &=&  - a\,h\,k\,n + b\,f\,j\,k + b\,g\,h\,k + b\,g\,h\,n 
\nnnl&&
- b\,h\,j\,m + b\,h\,k\,e 
- b\,k^2 e + c\,g\,h\,n
\nnnl&&
+ d\,f\,h\,n - f\,h\,j\,l - g\,h^2 l + h\,k\,l\,e  \scs
\nnnl[0.75mm]
\bar b_4\,&=&  - 4\,a\,f\,g\,h\,n + 2\,a\,h\,k\,n\,r + 4\,b\,f^2 g\,n 
- 4\,b\,f\,g\,k\,e
\nnnl&& 
+ 2\,b\,f\,g\,n\,e - b\,f\,j\,n\,r 
- 2\,b\,f\,k\,n\,r + 2\,b\,g\,h\,m\,e 
\nnnl &&
- b\,g\,h\,n\,r 
- 2\,b\,g\,k\,e^2  - b\,h\,k\,r\,e + b\,j\,k\,r\,e 
\nnnl&&
+ 2\,b\,k^2 \,r\,e
 + 4\,f\,g\,h\,l\,e - 2\,h\,k\,l\,r\,e  \scs
\nnnl[0.75mm]
\bar b^5_1 &=& a\,f\,h\,j - a\,g\,h^2 - b\,f^2 j + b\,f\,g\,h 
+ b\,f\,k\,e + 2\,b\,g\,h\,e 
\nnnl &&
- b\,h\,j\,r - b\,h\,k\,r 
+ c\,g\,h\,e - c\,h\,j\,r - d\,f\,h\,e + d\,h^2 r  \scs
\nnnl[0.75mm]
\bar b^5_2 &=& a\,h^2 k - a\,h\,j\,k - b\,f\,h\,k + b\,f\,j\,k 
- b\,g\,h\,n 
\nnnl &&
+ b\,h\,j\,m + b\,h\,k\,m - b\,h\,k\,e 
- b\,k^2 e + c\,h\,j\,m 
\nnnl &&
- c\,h\,k\,e - d\,h^2 m + d\,h\,k\,e  \scs
\nnnl[0.75mm]
\bar b^5_4 &=& 2\,f\,g\,n\,e - f\,j\,n\,r + 2\,g\,h\,m\,e 
- g\,h\,n\,r 
\nnnl&&
- 2\,g\,k\,e^2 - h\,k\,r\,e + j\,k\,r\,e  \scs
\nnnl[0.75mm]
\bar c_1 &=& a\,f\,j\,n + a\,g\,h\,n - b\,f\,j\,m + b\,f\,k\,e
\nnnl &&
 - b\,g\,h\,m + b\,g\,k\,e - c\,g\,n\,e - d\,f\,n\,e  \scs
\nnnl[0.75mm]
\bar c_2 &=& a\,k\,n^2 - 2\,b\,f\,j\,k + b\,f\,k\,n - 2\,b\,g\,h\,k 
\nnnl &&
+ b\,g\,k\,n + 2\,b\,k^2 e 
- b\,k\,m\,n - c\,g\,n^2 
\nnnl &&
- d\,f\,n^2 + f\,j\,l\,n + g\,h\,l\,n - k\,l\,n\,e  \scs
\nnnl[0.75mm]
\bar c_4 &=& 2\,a\,f\,g\,n^2 - a\,k\,n^2 r + 4\,b\,f\,g\,k\,e 
- 2\,b\,f\,g\,m\,n 
\nnnl &&
- 2\,b\,k^2 r\,e + b\,k\,m\,n\,r 
- 2\,f\,g\,l\,n\,e + k\,l\,n\,r\,e  \scs
\nnnl[0.75mm]
\bar c^5_1 &=&  - a\,f\,j\,n + a\,g\,h\,n + b\,f\,j\,m 
- b\,f\,k\,e 
\nnnl &&
- b\,g\,h\,m + b\,g\,k\,e + b\,h\,k\,r - b\,j\,k\,r
\nnnl &&
- c\,g\,n\,e + c\,j\,n\,r + d\,f\,n\,e - d\,h\,n\,r  \scs
\nnnl[0.75mm]
\bar c^5_2 &=&  - a\,h\,k + a\,j\,k - c\,j\,m + c\,k\,e 
+ d\,h\,m - d\,k\,e  \scs
\nnnl[0.75mm]
\bar c^5_4 &=& 0  \fs
\eea

\setcounter{figure}{0}
\setcounter{table}{0}
\section[Axial form factors at order $p^6$]
{Axial form factors at order \boldmath{$p^6$}}\label{app:axial}

In this appendix, we give the explicit formulae for the next-to-leading order
corrections to the axial form factors $A_1$, $A_2$, and $A_4$, as 
written out formally in \eqref{Aistruct}.
\begin{figure}
\centering
\vskip 2mm
\includegraphics[width=0.98\linewidth]{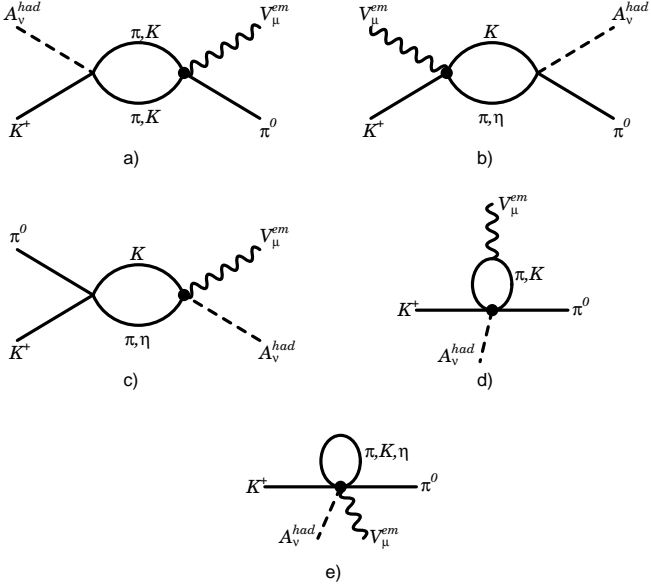}
\caption{Diagrams that contribute to the anomalous amplitude
  $A_{\mu\nu}$ at order $p^6$ [in a), the contribution from $\pi,\eta$
  intermediate states vanishes]. 
  Charges of the mesons running in the loops  are not indicated.  
  The filled vertices denote a contribution from the anomalous
  Lagrangian at order $p^4$. External line insertions in the tree
  diagram of order $p^4$ are not displayed.\label{fig:axialloops}}
\end{figure}
The necessary loop diagrams for this calculation are displayed in
Fig.~\ref{fig:axialloops}.
We find the following combinations of loop functions and
counterterms:
\begin{align}
S_1(s) &\eq    H_{\pi\pi}(s) +3 H_{K    K}(s) +\frac{16\pi^2}{3} \,C^r_{1s} \,s~,\notag\\
S_2(s) &\eq                                    \frac{16\pi^2}{3} \,C_{2s} \,s~,\notag\\
T_1(t) &\eq  \frac{1}{2} H_{K\pi}(t) + \frac{1}{2} H_{\eta K}(t) +\frac{16\pi^2}{3} \,C^r_{1t} \,t~,\notag\\
T_2(t) &\eq T_{K\pi}(t) - T_{\eta K}(t) +\frac{16\pi^2}{3} \,C^r_{2t} \,t ~,\notag\\
U_1(u) &\eq  \frac{1}{2} H_{K\pi}(u) + \frac{1}{2} H_{\eta K}(u) +\frac{16\pi^2}{3} \,C^r_{1u} \,u~,\notag\\
U_2(u) &\eq  2 H_{K  \pi}(u) +2 H_{\eta K}(u) +\frac{16\pi^2}{3} \,C^r_{2u} \,u~,\notag\\
X_1 &\eq             2 \mu_\pi -\mu_K -\mu_\eta 
  + \frac{16\pi^2}{3}\bigl(C^r_{1\pi} M_\pi^2 + C^r_{1K} M_K^2  \bigr)~, \notag\\
X_2 &\eq  \frac{19}{12}\mu_\pi -\frac{19}{6}\mu_K +\frac{1}{4}\mu_\eta  \no\notag\\
  & \hskip 0.5cm
  + \frac{16\pi^2}{3}\bigl(C^r_{2\pi} M_\pi^2 + C^r_{1K} M_K^2 \bigr) ~. 
\label{eq:axialstu}
\end{align}
The loop function $H_{ab}(x)$ is given by
\bea
H^r_{ab}(x) &=&
\frac{1}{12F^2}\biggl\{ \frac{\lambda(x,M_a^2,M_b^2)}{x}\bar{J}_{ab}(x) 
 + \frac{x-3\Sigma_{ab}}{24\pi^2} 
\label{Hdef} \\ 
&-& \frac{x}{32\pi^2}\log\frac{M_a^2M_b^2}{\mu^4} 
 - \frac{x\,\Sigma_{ab}-8M_a^2M_b^2}{32\pi^2\Delta_{ab}}\log\frac{M_a^2}{M_b^2} 
\biggr\} ~, \notag
\eea
where the loop function $\bar{J}_{ab}(x)$ is defined according to
\bea
\bar{J}_{ab}(t) &\eq& J_{ab}(t) - J_{ab}(0) ~,\label{defJbar}\\
J_{ab}(q^2) &\eq&
\frac{1}{i}\int \frac{d^dl}{(2\pi)^d}  
  \frac{1}{\bigl(M_a^2-l^2\bigr)\bigl(M_b^2-(l-q)^2\bigr)} ~. \notag
\eea
The other functions can also be written in relatively compact forms:
\begin{align}
T_{K\pi}^r(t) &= -\frac{1}{24F^2} \biggl\{ 
13t \biggl[ \bar{J}_{K\pi}(t) 
\no\\ & \hskip 9mm 
-\frac{1}{32\pi^2}\Bigl(\log\frac{\mk\mpi}{\mu^4}
+ \frac{\Sigma_{K\pi}}{\Delta_{K\pi}} \log\frac{\mk}{\mpi} \Bigl) \biggr]
\no\\ & \hskip 0.5cm
-
\biggl[
2\Sigma_{K\pi}-8\Delta_{K\pi} 
\no\\ & \hskip 9mm
+\!\Bigl(16\Sigma_{K\pi}+11\Delta_{K\pi}
  -\frac{8\Delta_{K\pi}^2}{t}\Bigr)\frac{\Delta_{K\pi}}{t} \biggr] 
\!\bar{J}_{K\pi}(t) 
\no\\ & \hskip 0.5cm
+\frac{M_K^2M_\pi^2(2\Delta_{K\pi}+t)}{4\pi^2t\Delta_{K\pi}}\log\frac{\mk}{\mpi}
\no\\ & \hskip 0.5cm
- \frac{(t-3\Sigma_{K\pi})(t-\Delta_{K\pi})}{12\pi^2t} \biggr \}~,  \\
T_{\eta K}^r(t)  &= \frac{1}{24F^2} \biggl\{ 
t \biggl[ \bar{J}_{\eta K}(t) 
\no\\ & \hskip 9mm
-\frac{1}{32\pi^2}\Bigl(\log\frac{\me\mk}{\mu^4}
+ \frac{\Sigma_{\eta K}}{\Delta_{\eta K}} \log\frac{\me}{\mk} \Bigl) \biggr]
\no\\ & \hskip 0.5cm
+\biggl[
2\Sigma_{\eta K}+8\Delta_{\eta K}
\no\\ & \hskip 9mm
-\Bigl(\frac{8}{3}\Sigma_{\eta K}+9\Delta_{\eta K}-\frac{8\Delta_{\eta K}^2}{t}\Bigr)\frac{\Delta_{K\pi}}{t}
 \biggr] 
\bar{J}_{\eta K}(t) 
\no\\ & \hskip 0.5cm
+\frac{\me\mk(2\Delta_{K\pi}-t)}{4\pi^2t\Delta_{\eta K}}\log\frac{\me}{\mk}
\no\\ & \hskip 0.5cm
- \frac{(t-3\Sigma_{\eta K})(t-\Delta_{K\pi})}{12\pi^2t} \biggr \}~,  \\
\mu_a &= \frac{M_a^2}{32\pi^2F^2}\log\frac{M_a^2}{\mu^2} ~,
\end{align}
where we have used $\Sigma_{ab}=M_a^2+M_b^2$, $\Delta_{ab}=M_a^2-M_b^2$,
and the two point function $\bar{J}_{ab}(x)$ as defined in \eqref{defJbar}.
The combinations of low-energy constants occurring in
\eqref{Aistruct} and \eqref{eq:axialstu} are given in Table~\ref{tab:coeff_cis}
according to the numbering in Ref.~\cite{p6anom}.
\begin{table}
\caption{The coefficients from \eqref{eq:axialstu} in terms of the
  renormalized low-energy constants $C_{i}^{Wr}$. For example,
  $C_{1s}^r=-C_{13}^{Wr}-5 C_{14}^{Wr}+\ldots$. 
  Constants without superscript $r$ are scale independent.
  \label{tab:coeff_cis}
}
\small{
\bea
{
\renewcommand{\arraystretch}{1.4}
\begin{array}{lrrrrrrrrrrr}
\hline
&
C_{1s}^r   &
C_{1t}^r   &
C_{1u}^r   &
C_{1\pi}^r &
C_{1K}^r   &
C_{2s}     &
C_{2t}^r   &
C_{2u}^r   &
C_{2\pi}^r &
C_{2K}^r   &
C_{4A} \\
\hline
C_{2}^{Wr}  &  0 &  0 & 0 & 24&-24 & 0 & 0 & 0 & 0 & 0 & 0
\\
C_{4}^{Wr}  &  0 &  0 & 0 & 16& 4  & 0 & 0 & 0 & 64& 0 & 0
\\
C_{5}^{Wr}  &  0 &  0 & 0 &-4 & 8  & 0 & 0 & 0 &-16& 0 & 0
\\
C_{7}^{W}   &  0 &  0 & 0 & 0 & 0  & 0 & 0 & 0 & 64&-16& 0
\\
C_{9}^{W}   &  0 &  0 & 0 & 0 & 0  & 0 & 0 & 0 &-48& 0 & 0
\\
C_{11}^{Wr} &  0 &  0 & 0 & 0 & 0  & 0 & 0 & 0 & 96&-48& 0
\\
C_{13}^{Wr} & -1 & -4 &-1 &-2 & 4  &-4 &-2 & 0 &-22& 2 &-16
\\
C_{14}^{Wr} & -5 &  7 &-2 &-6 &-3  &-20& 8 & 0 &-4 &-8 & 48
\\
C_{15}^{Wr} &  4 & -2 & 4 &-6 &-6  & 16& 8 & 0 &-40&-8 &-32
\\
C_{19}^{Wr} &  1 &  1 & 1 &-1 &-1  & 4 & 2 & 0 &-2 &-2 & 0
\\
C_{20}^{Wr} & -1 & -1 &-4 & 4 & 1  &-4 &-8 & 0 & 20& 8 & 16
\\
C_{21}^{Wr} & -4 & -4 &-4 & 4 & 4  &-16&-8 & 0 & 8 & 8 & 0
\\
C_{22}^{Wr} & 5/2& 5/2& 1 &-1 &-5/2&-6 & 2 & 4 & 4 &-2 & 8
\\
C_{23}^{W}  &-9/2&-9/2&-6 & 6 & 9/2&-6 &-6 & 0 & 12& 6 & 8
\\ \hline
\end{array}
}
\renewcommand{\arraystretch}{1.0}
\nonumber
\eea
}
\end{table}

\setcounter{figure}{0}
\setcounter{table}{0}
\section{Numerical parameters}\label{app:numerical}

\begin{sloppypar}
We denote the neutral pion and
charged kaon masses with $M_\pi$ and $M_K$,
respectively.
In numerical evaluations, we use
\begin{align}
M_K &\eq 493.68\MeV ~, &
M_{\pi} &\eq M_{\pi^0} \eq 134.98\MeV ~, \no \\
M_{\pi^\pm} &\eq 139.57\MeV ~, &
m_e &\eq 0.511\MeV ~, \\
F_\pi &\eq 92.4\MeV ~, &
F_K &\eq 1.22\,F_\pi ~. \no
\end{align} 
The $K_{e3}$ form factor is parameterized by
\beq
f_+(t) \eq f_+(0)\left[1+\lambda_+\frac{t}{M_{\pi^\pm}^2}
+\lambda''_+\frac{t^2}{M_{\pi^\pm}^4}+\cdots\right] ~.
\eeq
As explained in the main text, the precise values 
of $f_+(0)$ and  $\lambda_+$ do not matter in the present context.
For numerical evaluations, we use the parameter-free 
one-loop result including isospin corrections~\cite{mesonff}
\beq
f_+^{K^+\pi^0}(0) \eq 1.022\times f_+^{K^0\pi^-}(0)
\eq 0.998 ~.
\eeq
We stick to the low-energy constants chosen in Ref.~\cite{GKPV},
\beq
L_9^r(M_\rho) \eq 6.3\times 10^{-3} ~,~~ 
L_{10}^r(M_\rho)= -4.9\times 10^{-3} ~,
\eeq
which lead to a central value for $\lambda_+$ at one loop of
\beq
\lambda_+^c \eq 0.0294 ~.
\eeq
We express the low-energy constants in the following, scale-independent
form~\cite{BEG}: 
\beq\begin{split}
\bar{L}_9    &\eq L^r_9   (\mu) - \frac{1}{512\pi^2}\log\frac{\mpi M_K^4\me}{\mu^8}~, \\
\bar{L}_{10} &\eq L^r_{10}(\mu) + \frac{1}{512\pi^2}\log\frac{\mpi M_K^4\me}{\mu^8}~. 
\end{split}\label{barLi}\eeq
\end{sloppypar}

\end{appendix}



\end{document}